\documentclass[11pt,a4paper,notitlepage]{article}
\pdfoutput=1

\usepackage{jcappub}

\usepackage{lmodern}
\usepackage[T1]{fontenc}
\usepackage{graphicx}
\usepackage{enumitem}
\usepackage{amsmath}
\usepackage{booktabs}
\usepackage{array}
\usepackage{color}
\usepackage{amssymb}
\usepackage{mathrsfs}
\usepackage[T1]{fontenc}
\usepackage[latin1]{inputenc}
\usepackage[table]{xcolor}  
\usepackage{verbatim}
\usepackage{rotating}
\usepackage{afterpage}
\usepackage{dcolumn}
\usepackage{bm}
\usepackage[labelfont=sf,font+=small,format=hang,justification=RaggedRight]{subcaption}
\renewcommand{\d}{\mathrm{d}}
\newcommand{\e}[1]{\mathrm{e}^{{#1}}}

\newcommand{\vect}[1]{\bm{\mathrm{{#1}}}}
\newcommand{\im}{\mathrm{i}}

\newcommand{\ipleft}{\langle\kern-0.2em\langle}
\newcommand{\ipright}{\rangle\kern-0.2em\rangle}

\newcommand{\fNL}{f_{\mathrm{NL}}}
\newcommand{\fNLlocal}{\fNL^\text{local}}

\newcommand{\hatfNL}{\hat{f}_{\mathrm{NL}}}
\newcommand{\hatfNLlocal}{\hatfNL^\text{local}}
\newcommand{\hatfNLequi}{\hatfNL^\text{equi}}
\newcommand{\hatfNLortho}{\hatfNL^\text{ortho}}

\newcommand{\gNL}{g_{\mathrm{NL}}}
\newcommand{\tauNL}{\tau_{\mathrm{NL}}}

\newcommand{\hattauNL}{\hat{\tau}_{\mathrm{NL}}}

\newcommand{\nfNL}{n_{\fNL}}
\newcommand{\ntauNL}{n_{\tauNL}}

\newcommand{\Mp}{M_{\mathrm{P}}}

\newcommand{\vectkL}{\vect{k}_{\mathrm{L}}}
\newcommand{\kL}{k_{\mathrm{L}}}

\newcommand{\dimlessP}{\mathcal{P}}

\newcommand{\isomode}{\sigma}

\newcommand{\isocrit}{\isomode_{\text{c}}}
\newcommand{\isograd}{\isomode_{\text{step}}}
\newcommand{\Ngrad}{N_{\text{step}}}

\newcommand{\xplus}{\vect{x}_+}
\newcommand{\xls}{x_{\text{ls}}}

\newcommand{\kstar}{k_{\ell=1}}

\renewcommand{\leq}{\leqslant}
\renewcommand{\geq}{\geqslant}

\DeclareMathOperator{\Or}{O}

\newcommand{\semibold}[1]{{\fontseries{b}\selectfont{#1}}}
\newcommand{\para}[1]{\par\vspace{2mm}\noindent\semibold{{#1.}---}\ignorespaces}

\newcolumntype{s}{>{$\displaystyle}l<{$}}
\newcolumntype{t}{>{$\displaystyle}c<{$}}
\newcolumntype{u}{>{$\displaystyle}r<{$}}
\newcolumntype{v}{>{$\displaystyle}m{4cm}<{$}}

\newcolumntype{d}{D{!}{\;\pm\;}{-1}}

\title{\fontseries{s}\selectfont The hemispherical asymmetry from a scale-dependent inflationary bispectrum
\vspace{5mm}\hrule}
\author{\fontseries{s}\selectfont Christian T. Byrnes,}
\author{Donough Regan,}
\author{David Seery \\ and}
\author{Ewan R. M. Tarrant}
\affiliation{\small
Astronomy Centre, University of Sussex,
Falmer, Brighton BN1 9QH, UK}
\emailAdd{C.Byrnes@sussex.ac.uk}
\emailAdd{D.Regan@sussex.ac.uk}
\emailAdd{D.Seery@sussex.ac.uk}
\abstract{If the primordial
bispectrum is sufficiently large
then the CMB hemispherical power asymmetry
may be explained by
a large-scale mode of exceptional amplitude which perturbs
the $\zeta$ two-point function.
We extend previous calculations,
which were restricted to one- or two-source scenarios,
by providing a
method to compute the response of the two-point
function in any model yielding a `local-like' bispectrum.
In general, this shows that
it is not
the reduced bispectrum
$\fNL(k_1, k_2, k_3)$
which sources the amplitude and scale-dependence of the
mode coupling but rather a combination of `response functions'.
We discuss why it is difficult
to construct successful scenarios and enumerate
the fine-tunings which seem to be required.
Finally, we exhibit a concrete model
which can be contrived to match
the observational constraints
and show that
to a Planck-like experiment it would appear to have
$|\hatfNLlocal| \sim |\hatfNLequi| \sim |\hatfNLortho| \lesssim 1$.
Therefore, contrary to previous analyses,
we conclude that it is possible to generate
the asymmetry while respecting observational
constraints on the bispectrum and low-$\ell$ multipoles
even without tuning our location on the long-wavelength mode.}

\begin{document}	
\maketitle

\section{Introduction}
\label{sec:introduction}

Observations of the cosmic microwave background anisotropy on very large scales
have accumulated modest evidence for a small number of anomalies
in tension with our simplest picture of the early universe~\cite{Schwarz:2015cma}.
One of these is a roughly dipolar modulation of power
which enhances the temperature fluctuations in one hemisphere.
It is as if the power spectrum
$\dimlessP(k)$ of modes contributing to the CMB anisotropy
took the form
\begin{equation}
	\dimlessP^{\text{obs}}(k)
	\approx
	\frac{k^3 P(k)}{2\pi^2}
	\Big(
		1 + 2 A(k) \hat{\vect{p}} \cdot \hat{\vect{n}} + \cdots
	\Big) ,
	\label{eq:Pk-modulation}
\end{equation}
where $P(k)$ is the power spectrum for a statistically
homogeneous and isotropic
curvature perturbation synthesized by the early
universe.
The function $A(k)$ represents the scale-dependent amplitude of modulation,
and $\hat{\vect{p}}$ and $\hat{\vect{n}}$ are unit vectors in
the direction of maximum
asymmetry and the line of sight, respectively, measured from Earth.

This effect has been observed with amplitude
$A \approx 0.07$
on the largest scales in the
WMAP~\cite{Eriksen:2003db,Hansen:2004vq,Eriksen:2007pc,Hoftuft:2009rq}
and Planck~\cite{Ade:2013nlj,Ade:2015hxq}
microwave background surveys.
(See also Refs.~\cite{Paci:2013gs,Akrami:2014eta}.)
Numerically
this should be interpreted as an average amplitude
over a range of $k$ contributing to the low
multipoles of the CMB.
On smaller scales
Flender \& Hotchkiss obtained the constraint
$A \lesssim 0.0045$
suggesting that the modulation must exhibit a strong
scale dependence~\cite{Flender:2013jja},
and
improving an earlier constraint due to Hirata~\cite{Hirata:2009ar}.
More recently
Aiola et al.
succeeded in estimating $A$ as a function
of scale,
finding an approximate
fit to a power law $k^n$ with spectral index
$n$ of roughly
$-0.5$~\cite{Aiola:2015rqa}.%
	\footnote{Aiola et al. reported their constraint as an $\ell$-dependent modulation
	$A(\ell)$ of the angular power spectrum $C_\ell$.
	A power-law primordial spectrum of the form $P(k) \sim k^{-(3+n)}$
	induces an angular power spectrum of the form $C_\ell \sim \ell^{-(2+n)}$.
	Therefore the observed modulation $A(\ell)$ implies a primordial power
	spectrum modulated by the same power law.
	This result was noted recently by
	Adhikari, Shandera \& Erickcek~\cite{Adhikari:2015yya}.}
These measurements are expected to improve in the near future
owing to the arrival of new data from
polarization and CMB lensing~\cite{Mukherjee:2015wra}
which will extend the number of independent measurements of the
longest visible modes.

If this effect is real, its implications for early universe models are not
yet clear.
Erickcek, Kamionkowski \& Carroll suggested that a modulation of this kind
could be produced by a
fluctuation
with wavelength longer than the scale of the last-scattering surface
and an anomalously large amplitude~\cite{Erickcek:2008sm,Erickcek:2008jp}.
The linear gradient
generated by this mode
could produce a suitable asymmetry
if the small-scale $\zeta$ two-point function 
responds to its presence.
Any such response implies a nontrivial
correlation
between the long wavelength fluctuation
$\delta \isomode(\vectkL)$
and two curvature modes
$\zeta(\vect{k})$ with $k \gg \kL$,
and therefore should be controlled by
`squeezed' momentum configurations of the three-point function
$\langle \delta \isomode(\vect{k}_3) \zeta(\vect{k}_1) \zeta(\vect{k}_2) \rangle$
with $k_3 \ll k_1$, $k_2$.
But
this three-point function will also contribute
to the observable three-point function
$\langle \zeta(\vect{k}_3) \zeta(\vect{k}_1) \zeta(\vect{k}_2) \rangle$
at a level which depends on the way in which
$\delta \isomode$ contributes to $\zeta$.
Therefore, in this scenario, we should expect a modulation of the
form~\eqref{eq:Pk-modulation}
to be accompanied by some level of non-Gaussianity---although its precise
amplitude cannot be predicted without further information.

Efforts have been made to quantify this effect.
Working with a `single-source' model in which
the fluctuations of a single field dominate $\zeta$,
Lyth used the separate universe picture to compute how
short-scale $\zeta$ modes would be biased by a long-wavelength
perturbation~\cite{Lyth:2013vha,Lyth:2014mga}.
This calculation suggested
that the response of the $\zeta$ two-point function would be
characterized by the amplitude of the reduced
bispectrum, $\fNL$,
and that compatibility with constraints on the CMB quadrupole
would
probably require $|\fNL| \gtrsim 10$ or more.

If this estimate were applicable to the amplitude of
three-point correlations in squeezed configurations
then it would be incompatible
with the simplest slow-roll, single-field inflationary models.
Therefore
achieving sufficient modulation would entail more complex dynamics,
either during or after inflation.
Even in these models, an amplitude of order
$|\fNL| \gtrsim 10$
would place the scenario in tension with
recent Planck constraints on the amplitude of
three-point correlations in squeezed configurations.

Similar conclusions were reached by
Abolhasani, Bagram, Namjoo
\& Firouzjahi~\cite{Namjoo:2013fka,Abolhasani:2013vaa,Namjoo:2014nra},
Kanno, Sasaki \& Tanaka~\cite{Kanno:2013ohv}
and
later
Kobayashi,
Cort\^{e}s \& Liddle~\cite{Kobayashi:2015qma}.
A variant of the scenario, based on the response being generated
by a large trispectrum rather than bispectrum,
was suggested by Kenton, Mulryne \& Thomas~\cite{Kenton:2015jga}.
But none of these analyses
explicitly accounted
for the strong scale-dependence
of three-point correlations
which is presumably entailed by the strong scaling
of $A(k)$.
At a minimum this would require specification of the
scale at which any amplitude constraint was intended to apply.
Moreover the separate universe approach used in these
investigations does
not make manifest which momentum configurations
of the three-point function are being invoked.
This is not merely a pedantic point.
If strong scaling is present the amplitude of
correlation on squeezed configurations can
differ by orders of magnitude
compared to
the amplitude on equilateral configurations---even for
configurations
of the same overall scale.%
	\footnote{Here (and in the remainder of this paper)
	we are taking the \emph{scale} of a momentum
	configuration
	specified by the triangle $\vect{k}_1 + \vect{k}_2 + \vect{k}_3 = 0$
	to mean its perimeter
	$k_t = k_1 + k_2 + k_3$.
	Its \emph{shape} is specified by the ratios $k_i/k_t$ of its
	sides to $k_t$.
	For squeezed configurations one side becomes much smaller than the
	other two. Taking this side to be $k_3$ that gives $k_3 / k_t \ll 1$
	while the other two sides have roughly $k_1/k_t \sim k_2/k_t \sim 0.5$.}
For example, in Ref.~\cite{Lyth:2013vha}
it was suggested that the amplitude relevant
for determining the response of the $\zeta$ two-point function
would come from equilateral configurations.
The amplitude in squeezed configurations might then be very different.
Since it is the squeezed configurations which mostly contribute
to present observational constraints,
for example through their contribution to CMB estimators
for the amplitude
$\fNLlocal$
of the `local' template,
this could dramatically change our conclusions regarding the
viability of the model or even the requirement for
dynamics beyond slow-roll, single-field inflation.

\para{Summary}%
In this paper we have three major goals.

First, we
explain how to compute the response of the
$\zeta$ two-point function to long-wavelength perturbations in
scenarios more general than the single-source model.
Our approach shows explicitly how the response depends on
information embedded in the squeezed limit of the
three-point function
$\langle \delta \isomode(\vect{k}_3) \zeta(\vect{k}_1) \zeta(\vect{k}_2) \rangle$,
and clarifies
which momentum configurations are relevant.
It does not rely on the separate universe method,
although in some circumstances that could be used to compute
the required correlation functions.

Second, despite the effort
which has been invested in studying the Erickcek et al. scenario,
it is still unclear what is entailed in an inflationary model
(perhaps extended by a later curvaton phase)
that achieves a suitable asymmetry by this
mechanism.
Explicit models were studied by
McDonald~\cite{McDonald:2013qca}
and
Kanno et al.~\cite{Kanno:2013ohv};
but
if observation forces us to consider
early-universe scenarios for the origin of the asymmetry---%
and this is not yet clear---%
we should like qualitative guidance concerning
the features to be expected.
We explain in general terms why it is difficult to manufacture
a scenario which produces a bispectrum with suitable
scale-dependence without simultaneously
producing other undesirable features, such as an
unacceptably tilted power spectrum
or a large trispectrum.

Third, we address the issue of observational constraints.
As explained above,
the strong running with scale
entailed by the scale-dependence of $A(k)$
makes it unclear
how large a bispectrum amplitude is acceptable.
The estimates made in Refs.~\cite{Lyth:2013vha,Lyth:2014mga,Kanno:2013ohv,Kobayashi:2015qma}
suggest that the required amplitude may be too
large, but these cannot be compared directly
to constraints reported by Planck or WMAP.
To make a comparison we must determine how the amplitude
of the bispectrum varies with scale and squeezing.
We exhibit a concrete (but contrived) model
which satisfies current observational constraints.
The construction of this model
exemplifies the general difficulties encountered in building
a successful scenario---but
having done so,
we use it to demonstrate the shape and magnitude of the
three-point correlations which it produces.
We use these to estimate how a Planck-like
experiment would view the bispectrum
through the response of the estimators
for $\hatfNLlocal$, $\hatfNLequi$ and $\hatfNLortho$.
It is then possible to address the viability of the scenario.

\para{Structure}%
This paper is structured as follows.
In~\S\ref{sec:biasing} we develop
a formalism which can be used to compute how an arbitrary
long-wavelength mode biases the $\zeta$ two-point function.
The key results are collected in~\S\S\ref{sec:ope}--\ref{sec:perturbed-2pf}.
Our formalism applies to an arbitrary inflationary model
provided the squeezed limit of the bispectrum is not suppressed
in a sense to be made precise below.
It does not make use of the slow-roll approximation.
In~\S\ref{sec:single-source} we explain how to relate our
approach to the existing literature, including
Refs.~\cite{Lyth:2013vha,Lyth:2014mga,Kanno:2013ohv,Kobayashi:2015qma,Kenton:2015jga}.

In~\S\ref{sec:why-hard} we give a heuristic argument
explaining why it is difficult
to build inflationary models with the correct response,
even if we allow for effects subsequent to inflation such as
a curvaton era. In~\S\ref{sec:scale-dependence}
we explain what is required to produce a bispectrum with the
correct scaling properties.
In~\S\ref{sec:no-const-eta} we show that the simplest of these
scenarios,
where the scale dependence is
generated by a large negative $\eta$-parameter,
has difficulty generating a bispectrum of sufficient
amplitude
during the inflationary epoch.
We sketch the problems encountered if we attempt to go beyond
the simplest scenario.

\S\ref{sec:working-model}
describes a working model which produces a suitable response
during inflation
by introducing a sharp feature in the potential.
This evades the constraints discussed in~\S\ref{sec:why-hard},
but also exemplifies the tunings which seem required to construct
a viable model.
We outline the model in~\S\ref{sec:tanh-model}
and give numerical results for the
response of the two-point function to biasing.
In~\S\ref{sec:gz-effect}
we compute constraints on the amplitude
of the large-scale mode from
the Grischuk--Zel'dovich
effect,
and in~\S\ref{sec:bsp-shape}
we discuss constraints from the bispectrum
including estimates for the response of the Planck
estimators $\hatfNLlocal$, $\hatfNLequi$
and $\hatfNLortho$.
Finally, we summarize our conclusions in~\S\ref{sec:conclusions}.

\para{Notation}%
Throughout this paper
we adopt units in which $c=\hbar=1$.
The reduced Planck mass is defined by $\Mp^2 = (8\pi G)^{-1/2}$, where
$G$ is Newton's gravitational constant.
We work with a collection of light scalar fields
and their momenta,
initially indexed by Greek labels
$\{ \alpha$, $\beta, \ldots \}$.
In order to work with compact formulae we adapt this
summation convention when multiple
times of
evaluation are under discussion, as
described in~\S\ref{sec:scale-dependence}.

We collect our notational conventions in Table~\ref{table:notation}.

\begin{table}

  \begin{center}
    \small
	\heavyrulewidth=.08em
	\lightrulewidth=.05em
	\cmidrulewidth=.03em
	\belowrulesep=.65ex
	\belowbottomsep=0pt
	\aboverulesep=.4ex
	\abovetopsep=0pt
	\cmidrulesep=\doublerulesep
	\cmidrulekern=.5em
	\defaultaddspace=.5em
	\renewcommand{\arraystretch}{1.5}
        
    \rowcolors{2}{gray!25}{white}

    \begin{tabular}{sll}
    	\toprule
    	\text{\semibold{notation}} & \semibold{meaning} & \semibold{definition} \\
		P(k) & power spectrum for $\langle \zeta \zeta \rangle$ & Eq.~\eqref{eq:L-box-2pf} \\
      	\dimlessP(k) & dimensionless power spectrum of $\langle \zeta \zeta \rangle$ & Eq.~\eqref{eq:Pk-modulation} \\
		B(k_1, k_2, k_3) & bispectrum for $\langle \zeta \zeta \zeta \rangle$ & Eq.~\eqref{eq:zeta-bispectrum} \\
		T(\vect{k}_1, \vect{k}_2, \vect{k}_3, \vect{k}_4) & connected trispectrum for $\langle \zeta \zeta \zeta \zeta \rangle$ 
			& Eq.~\eqref{eq:zeta-trispectrum} \\
		\fNL(k_1, k_2, k_3) & reduced bispectrum for $\langle \zeta \zeta \zeta \rangle$ & Eq.~\eqref{eq:reduced-bispectrum} \\
		\tauNL(\vect{k}_1, \vect{k}_2, \vect{k}_3, \vect{k}_4) & amplitude of $\tauNL$-mode in $T$ & Eq.~\eqref{eq:tauNL-def} \\
      	\dimlessP_\isomode(k) & dimensionless power spectrum for $\langle \isomode \isomode \rangle$ & Eq.~\eqref{eq:dimlessP-sigma-def} \\
      	\dimlessP_{\zeta_\isomode}(k) & dimensionless power spectrum for
      		$\langle \zeta_\isomode \zeta_\isomode \rangle$, where
      		$\zeta_\isomode = N_\isomode \delta\isomode$ & Eq.~\eqref{eq:R-def} \\
      	\Sigma^{\alpha\beta} & power spectrum for $\langle \delta\phi^\alpha \delta\phi^\beta \rangle$
      		& Eq.~\eqref{eq:field-powerspectrum-def} \\
      	B^\lambda(k_1, k_2, k_3) & bispectrum for $\langle \delta\phi^\lambda \zeta \zeta \rangle$
      		& Eq.~\eqref{eq:mixed-bispectrum-def} \\
      	A(k) & amplitude of asymmetry & Eq.~\eqref{eq:Pk-modulation} \\
      	\rho_\mu(k) & linear response function for $\delta\phi^\mu$ & Eq.~\eqref{eq:response-function} \\
      	\rho_\zeta(k) & linear response function for $\zeta$ & Eq.~\eqref{eq:zeta-response} \\
      	E & enhancement or exceptionality of long-wavelength mode & Eq.~\eqref{eq:modulating-mode} \\
      	\alpha, \vectkL & scale of long-wavelength mode & Eq.~\eqref{eq:alpha-def} \\
      	\xls & comoving distance to last-scattering surface & p.~\pageref{def:xls} \\
      	\hat{\vect{n}} & orientation of line-of-sight from Earth & p.~\pageref{def:xls} \\
      	\hat{\vect{p}} & orientation of exceptional mode & Eq.~\eqref{eq:alpha-def} \\
      	R & relative contribution of $\isomode$ to $\dimlessP(k)$ & Eq.~\eqref{eq:R-def} \\
      	N_\alpha, N_{\alpha\beta} & gauge transformation from field fluctuations to $\zeta$ & Eq.~\eqref{eq:gauge-transformation} \\
      	\Gamma^\alpha_a, \Gamma^\alpha_{ab} & separate-universe coefficients,
      		depending on \emph{two} times & Eq.~\eqref{eq:separate-universe-fields} \\
      	{u^\alpha}_\beta, {u^\alpha}_{\beta\gamma} & transport equation coefficients
      		& Eqs.~\eqref{eq:gamma2-forward}, \eqref{eq:gamma3-forward} \\
      	\eta_\isomode & $\eta$-parameter $V_{\isomode\isomode}/3H^2$ for $\isomode$ & \\
      	\xi_\isomode & parameter $\Mp V_{\isomode\isomode\isomode}/3H^2$ for $\isomode$ & Eq.~\eqref{eq:xi-equation} \\
      	k_t & perimeter of momentum $n$-gon & p.~\pageref{def:kt} \\
      	\kstar & roughly corresponds to $\ell=1$, $\kstar \approx 1/14,000 \, \text{Mpc}^{-1}$
      		& Eq.~\eqref{eq:tanh-response} \\
      	\bottomrule
    \end{tabular}
  \end{center}
  
  \caption{Notation used in this paper. Time-dependent
  quantities such as $n$-point functions,
  power spectra and bispectra
  are always evaluated at the time of interest unless otherwise
  specified.
  Note that this includes the gauge-transformation
  coefficients $N_\alpha$, $N_{\alpha\beta}$.
  \label{table:notation}}
  
\end{table}

\section{Biasing the two-point function by a long-wavelength mode}
\label{sec:biasing}

In this section we obtain a formula for the response
of a short-wavelength two-point function to
the presence of an underlying
perturbation with much longer wavelength.
Our result will be
valid, up to gradient-suppressed corrections,
to linear order in the amplitude
of the long-wavelength mode
but non-perturbatively in the short-wavelength modes.
We work with the $\zeta$ two-point function because this is the case
to which we will eventually apply our result, but the method is
general and can be used to study the biasing of any $n$-point function.

\subsection{The operator product expansion}
\label{sec:ope}

\begin{figure}
	\begin{center}
		\includegraphics[scale=1.0]{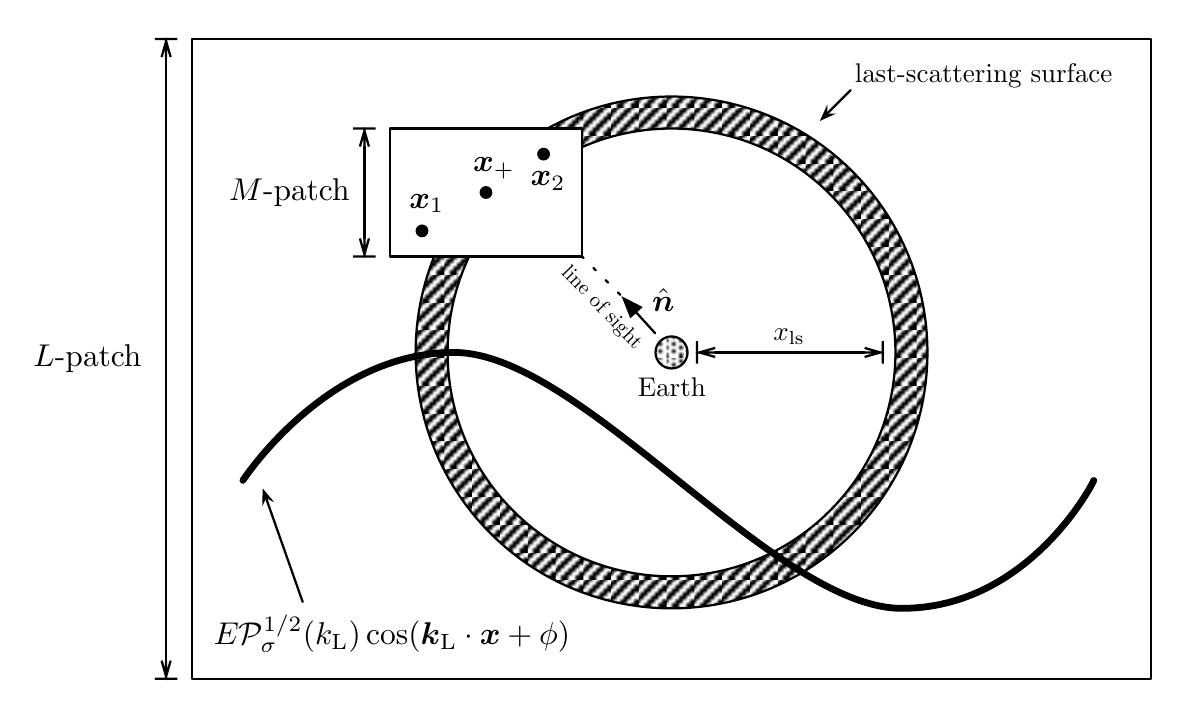}
	\end{center}
	\caption{Modulation of the power spectrum measured
	in an $M$-sized patch
	embedded within a larger $L$-sized patch.
	The $L$-patch is crossed by a long-wavelength mode~\eqref{eq:modulating-mode}
	whose amplitude is enhanced above the typical amplitude
	by a factor $E$.
	In applications, the $M$-patch [located at the aggregate
	coordinate $\xplus$; see Eq.~\eqref{eq:xplus}] would be centred on the last-scattering
	surface.
	\label{fig:LMpatch}}
\end{figure}

Consider a region of spacetime with comoving spatial extent $M$, within which
we wish to predict the $\zeta$ two-point function;
see Fig.~\ref{fig:LMpatch}.
We imagine that this region is enclosed within
a uniform larger patch of extent $L \gg M$,
and we suppose that an early-universe mechanism such as inflation has
seeded a set of statistically isotropic and homogeneous fluctuations within
this large patch.
In the scenario of Erickcek et al.,
the particular realization
within the $L$-patch
contains a rare large-amplitude mode with wavenumber $\kL \ll 1/M$.

Within the $M$-patch, the two-point function would respond to an infinitely
long wavelength perturbation
as if it were a shift in the zero-mode of the fields.
Therefore to linear order in the amplitude of the perturbation,
\begin{equation}
	\langle \zeta(\vect{x}_1) \zeta(\vect{x}_2) \rangle_M
	=
	\langle \zeta(\vect{x}_1) \zeta(\vect{x}_2) \rangle_L
	+
	\delta \phi^\mu
	\frac{\partial}{\partial \phi^\mu}
	\langle \zeta(\vect{x}_1) \zeta(\vect{x}_2) \rangle_L
	+ \cdots ,
	\label{eq:infinite-k-OPE}
\end{equation}
where `$\cdots$' denotes terms of second-order or higher
in $\delta\phi$ which we have neglected.
Except in~\S\ref{sec:why-hard}
we are not using the slow-roll approximation, so
$\delta \phi^\mu$ runs over the perturbations in
the scalar fields and their momenta.
The same is true for $\partial / \partial \phi^\mu$.

Our interest is in perturbations
with large but finite
wavelength.
For such perturbations~\eqref{eq:infinite-k-OPE}
represents the beginning of a series
describing the response
of $\langle \zeta \zeta \rangle$
to the position-dependent
fluctuation $\delta\phi^\mu(\xplus)$,
where $\xplus$ is the aggregate
position of the $M$-patch surrounding
$\vect{x}_1$ and
$\vect{x}_2$~\cite{Creminelli:2011rh,Pajer:2013ana,Hinterbichler:2012nm,Creminelli:2012ed},
\begin{equation}
	\xplus = \frac{\vect{x}_1 + \vect{x}_2}{2} .	
	\label{eq:xplus}
\end{equation}
The two-point function will respond not only to the displacement
$\delta\phi^\mu$
but also other local operators built from its gradients
such as $\partial^2 \delta\phi^\mu$.
Therefore
\begin{equation}
\begin{split}
	\langle \zeta(\vect{x}_1) \zeta(\vect{x}_2) \rangle_M
	=
	\mbox{}
	&
	\langle \zeta(\vect{x}_1) \zeta(\vect{x}_2) \rangle_L
	+
	\delta \phi^\mu(\xplus)
	\frac{\partial}{\partial \phi^\mu}
	\langle \zeta(\vect{x}_1) \zeta(\vect{x}_2)	\rangle_L
	\\
	&
	\mbox{}
	+ \partial^2 \delta\phi^\mu(\xplus)
	\frac{\partial}{\partial ( \partial^2 \phi^\mu )}
	\langle \zeta(\vect{x}_1) \zeta(\vect{x}_2) \rangle_L
	+ \cdots ,
	\label{eq:ope}
\end{split}
\end{equation}
where `$\cdots$' denotes contributions from other
local operators which we have not written explicitly.
In Eq.~\eqref{eq:ope} all quantities are evaluated at the same time.
A similar expansion was used by
Mirbabayi \& Simonovi\'{c}~\cite{Mirbabayi:2015hva}.

If $\delta\phi^\mu$ contains only long-wavelength contributions
then its gradients will be suppressed.
Where these suppressed terms can be neglected the
second term in Eq.~\eqref{eq:ope}
will furnish the dominant response
and the two-point function
reacts
to all long-wavelength modes in nearly the same way.
But if
$(\partial / \partial \phi^\mu) \langle \zeta \zeta \rangle$
is small or zero
then the leading correction may come instead from
a term such as $\partial^2 \delta\phi^\mu$.
In these cases the two-point function responds
differently
to perturbations of different wavelengths.

In this paper we focus on
scenarios in which
the dominant long-wavelength response comes from the
$\delta\phi^\mu$ operator.
This includes most models of interest
because if the dominant response involves gradients
it will be suppressed by powers of the
ratio $\kL/k$, where $k$ is a typical wavenumber
contributing to $\langle \zeta \zeta \rangle$
and $\kL$ is a long-wavelength mode in
$\delta\phi^\mu$.
We will see later that this suppression would make it
difficult to generate a suitable asymmetry without
a bispectrum of very large amplitude.
In addition the scale-dependence may be
incorrect; see Ref.~\cite{Adhikari:2015yya}.

The $\delta\phi^\mu$
operator in Eq.~\eqref{eq:ope} was used by Schmidt \& Hui to
compute the linear response of the two-point function
to immersion within a bath of
long-wavelength modes~\cite{Schmidt:2012ky}.%
	\footnote{Schmidt \& Hui's scenario is very similar to the
	one considered in this paper, except that we will take
	a single long-wavelength mode
	to have an exceptional amplitude.
	In Schmidt \& Hui's scenario all the long-wavelength modes have
	a typical amplitude
	and the resulting anisotropy
	receives contributions from all of them.
	What is observed in a typical region
	with scale comparable to the horizon at the time
	of last scattering
	therefore depends
	on the distribution of large-scale modes,
	which was studied by Adhikari et al.~\cite{Adhikari:2015yya}.
	They found the tail to be sufficiently
	broad that an exceptional amplitude might not be
	required
	if the bispectrum amplitude is sufficiently large
	on the largest observable scales.
	See also the discussion in~\S\ref{sec:bsp-shape} and~\S\ref{sec:conclusions}.}
Their scenario was developed by
Adhikari, Shandera \& Erickcek~\cite{Adhikari:2015yya}.
Later Pajer, Schmidt
\& Zaldarriaga~\cite{Pajer:2013ana}
demonstrated that the precise linear combination of
$\vect{x}_1$ and $\vect{x}_2$
appearing in Eq.~\eqref{eq:xplus}
is immaterial in the squeezed limit up to
corrections of order $\Or(k^2)$.
The discussion given here is closest to that of
Namjoo et al.~\cite{Namjoo:2013fka}
and
Kenton \& Mulryne~\cite{Kenton:2015lxa},
both of whom effectively used~\eqref{eq:ope}
to study biasing of a three-point function by
a long wavelength mode.
A similar method has been proposed by
Chiang et al. to measure the squeezed limit of
the primordial bispectrum using large-scale structure~\cite{Chiang:2014oga}.

\para{Response function}%
To use Eq.~\eqref{eq:ope}
we discard all gradient-suppressed
terms and take its Fourier transform within
the $M$-patch, leaving the location $\xplus$
of this patch fixed.
If the power spectrum $P(k)$ within the $L$-patch
satisfies
\begin{equation}
	\langle
		\zeta(\vect{k}_1)
		\zeta(\vect{k}_2)
	\rangle_L
	=
	(2\pi)^3
	\delta(\vect{k}_1 + \vect{k}_2)
	P(k) ,
	\label{eq:L-box-2pf}
\end{equation}
where $k = |\vect{k}_1| = |\vect{k}_2|$ is the common magnitude
of $\vect{k}_1$ and $\vect{k}_2$,
the result can be written
\begin{equation}
	\langle
		\zeta(\vect{k}_1)
		\zeta(\vect{k}_2)
	\rangle_M
	=
	(2\pi)^3
	\delta(\vect{k}_1 + \vect{k}_2)
	P(k)
	\Big(
		1
		+ \delta\phi^\mu(\xplus) \rho_\mu(k)
		+ \cdots
	\Big) .
	\label{eq:twopf-response}
\end{equation}
In Eq.~\eqref{eq:twopf-response}
we have introduced the
\emph{linear response function}
$\rho_\mu(k)$,
defined to be a rescaled derivative of $P(k)$
with respect to the zero-modes within the $L$-patch,
\begin{equation}
	\rho_\mu(k)
	\equiv
	\frac{1}{P(k)}
	\frac{\partial P(k)}{\partial \phi^\mu}
	.
	\label{eq:response-function}
\end{equation}
It is a function only of $k$, and as we have explained
it does not depend
on the wavenumber of the source.
If we require terms of higher-order in
the amplitude of the long-wavelength
mode then~\eqref{eq:twopf-response}
could be extended
by defining
higher-order response functions
$\rho_{\mu\nu} = P(k)^{-1} \partial_\mu\partial_\nu P(k)$,
$\rho_{\mu\nu\lambda} = P(k)^{-1} \partial_\mu\partial_\nu\partial_\lambda P(k)$,
and so on,
which measure the quadratic, cubic
or higher terms in the expansion.

Once we have obtained $\rho_\mu(k)$
it is sufficient (if nonzero)
to characterize linear biasing within the $M$-patch.
We can compute it by any convenient method---for example,
by constructing a numerical derivative,
or even by analytic differentiation if a closed-form expression
can be found.
However, as explained in~{\S\ref{sec:introduction}},
we expect that
the response of the two-point function should be controlled
by squeezed configurations of the
three-point function
$\langle \delta \phi^\mu \zeta \zeta \rangle$.
To determine this relationship
we return to~\eqref{eq:ope},
but
now regarded as an operator product expansion
for the bilinear
$\zeta(\vect{x}_1) \zeta(\vect{x}_2)$
within the $L$-patch.
Correlations between this bilinear
and a distant point $\vect{x}_3$
are controlled by the expansion.
If we arrange that $\langle \delta\phi^\mu(\vect{x}) \rangle_L = 0$
then Eq.~\eqref{eq:ope}
asserts that the leading contribution comes from modulation of the
two-point function,
\begin{equation}
	\langle \delta\phi^\lambda(\vect{x}_3) \zeta(\vect{x}_1) \zeta(\vect{x}_2) \rangle_L
	\approx
	\langle \delta\phi^\lambda(\vect{x}_3) \delta\phi^\mu(\xplus) \rangle_L
	\frac{\partial}{\partial \phi^\mu} \langle \zeta(\vect{x}_1) \zeta(\vect{x}_2) \rangle_L
	\quad
	\text{if $|\vect{x}_3 - \xplus| \gg |\vect{x}_1 - \vect{x}_2|$}
	.
	\label{eq:distant-correlation}
\end{equation}
We now take the Fourier transform within the $L$-patch.
For $k_3$ much less than $k_1$, $k_2$ the dominant contribution
to the Fourier integral will come from spatial configurations
for which Eq.~\eqref{eq:distant-correlation}
describes the behaviour of the three-point function.
Defining the bispectrum of the mixed correlation function by
\begin{equation}
	\langle \delta\phi^\lambda(\vect{k}_3) \zeta(\vect{k}_1) \zeta(\vect{k}_2) \rangle_L
	=
	(2\pi)^3
	\delta(\vect{k}_1 + \vect{k}_2 + \vect{k}_3)
	B^\lambda(k_1, k_2, k_3)	
	\label{eq:mixed-bispectrum-def}
\end{equation}
it can be shown that
the Fourier transform yields
\begin{equation}
	B^\lambda(k_1, k_2, k_3)
	\approx
	\Sigma^{\lambda\mu}(k_3)
	\partial_\mu P(|\vect{k}_1 - \vect{k}_3/2|)
	\quad
	\text{if $k_3 \ll k_1, k_2$}
	,
	\label{eq:mixed-bispectrum-ope}
\end{equation}
where $\Sigma^{\alpha\beta}$ is the power spectrum of
the field fluctuations within the $L$-patch,
\begin{equation}
	\langle
		\delta\phi^\alpha(\vect{k}_1)
		\delta\phi^\beta(\vect{k}_2)
	\rangle_L
	=
	(2\pi)^3
	\delta(\vect{k}_1 + \vect{k}_2)
	\Sigma^{\alpha\beta}(k_1) .
	\label{eq:field-powerspectrum-def}
\end{equation}
Eq.~\eqref{eq:mixed-bispectrum-ope}
shows
that the response function can be extracted from
knowledge of $B^\lambda(k_1, k_2, k_3)$ and
$\Sigma^{\alpha\beta}(k)$.
To leading order in $k_3/k_1 \approx k_3/k_2$
the required combination is
\begin{equation}
	\rho_\mu(k)
	\approx
	\frac{1}{P(k)}
	[\Sigma^{-1}(k_3)]_{\mu\lambda}
	B^\lambda(k, k, k_3)
	\quad
	\text{if $k_3 \ll k$}
	.
	\label{eq:calculate-response}
\end{equation}
Provided the $\delta\phi^\mu$ operator dominates
the long-wavelength response,
the operator product expansion
guarantees that the right-hand side becomes
independent of $k_3$ for sufficiently squeezed
configurations.
In this case
$\rho_\mu(k)$ can be determined approximately
from any suitable configuration of this type.
The wavenumber $k$ represents the nearly
equal magnitude of the short modes,
$k \approx k_1 \approx k_2$.

Alternatively,
it may happen that~\eqref{eq:calculate-response}
does not approach a nonzero limit for small $k_3$.
This indicates that the $\delta\phi^\mu$ operator
in~\eqref{eq:ope} did not control
the response to long-wavelength perturbations,
and a higher-order operator or
one of the gradient-suppressed contributions
is instead dominant.
Examples of bispectra for which this occurs
include the equilateral and orthogonal
templates, because these diverge
more slowly than the power-spectrum $\Sigma$
as $k_3 \rightarrow 0$.
In such cases it is possible to modify the discussion
in this section to extract a corresponding response
function, but we will not pursue this possibility;
for the reasons explained above
the response is suppressed,
and a bispectrum of large amplitude is required to generate
a suitable asymmetry.
For an alternative approach to these templates see
Adhikari, Shandera \& Erickcek~\cite{Adhikari:2015yya}.

For the models to be studied in this paper Eq.~\eqref{eq:calculate-response}
is sufficient.
These have `local-like' bispectra
in the sense that
correlations in the squeezed limit
are not suppressed by powers of $\kL/k$, but there is no
requirement that the bispectrum shape is a close match
to the local template.
Indeed, to generate a suitable scale-dependent asymmetry $A(k)$
we will require
departures from the local shape.
In these models Eq.~\eqref{eq:calculate-response}
shows that the response of the two-point function
depends on the squeezed limit of the mixed
bispectrum $B^\lambda(k_1, k_2, k_3)$,
although because $\rho_\mu(k)$ is itself independent
of $k_3$ it cannot depend on ratios
such as $k_3/k_t \sim k_3/k$ which measure the
squeezing.
We also note that no part of our analysis required the
slow-roll approximation.

Finally,
if we wished to include the quadratic or cubic response
functions $\rho_{\mu\nu}$,
$\rho_{\mu\nu\lambda}$
then these could be estimated in a similar way,
by studying the
double-soft limit of
$\langle \delta\phi^\alpha \delta\phi^\beta \zeta \zeta \rangle$
or the
triple-soft limit of
$\langle \delta\phi^\alpha \delta\phi^\beta \delta\phi^\gamma \zeta \zeta \rangle$---%
in each case as the momenta associated with the field fluctuations
become much smaller than those carried by $\zeta$.

\para{OPE determines collapsed trispectrum}%
The operator product expansion determines
certain other $n$-point functions on a subset of configurations.
In particular, an analysis similar to that of
Eqs.~\eqref{eq:distant-correlation} and~\eqref{eq:mixed-bispectrum-ope}
shows
that Eq.~\eqref{eq:ope}
determines the
four-point function on `collapsed' configurations
in terms of the response functions,
\begin{equation}
	\langle
		\zeta(\vect{k}_1)
		\zeta(\vect{k}_2)
		\zeta(\vect{k}_3)
		\zeta(\vect{k}_4)
	\rangle_L^{\text{connected}}
	\approx
	(2\pi)^3 \delta\big( \sum_i \vect{k}_i \big)
	\Sigma^{\alpha\beta}(|\vect{k}_3 + \vect{k}_4|)
	\rho_\alpha(k_2) \rho_\beta(k_4)
	P(k_2) P(k_4)
	\label{eq:collapsed-trispectrum}
\end{equation}
if $|\vect{k}_3 + \vect{k}_4| \ll k_2$, $k_4$.
Further formulas can be found for increasingly
restrictive configurations of the
higher $n$-point functions, but we do not study these
in detail because there is no imminent
prospect of
restrictive observational
constraints.

\subsection{The perturbed two-point function}
\label{sec:perturbed-2pf}

We now estimate the perturbation
in the $M$-patch two-point function
$\langle \zeta(\vect{k}_1) \zeta(\vect{k}_2) \rangle_M$
produced by a collection of long-wavelength modes
crossing the $L$-patch.
In this paper we are principally interested in the case
where there is a rare fluctuation with enhanced amplitude
associated with a single wavenumber $\vectkL$,
and
for simplicity we will assume it is present in
just one species which we denote $\isomode$.
The case of multiple contributing species (or wavenumbers)
can be handled by obvious modifications of our formulae.

Neglecting correlations between species, the typical
variance
of fluctuations in $\isomode$
will be given by
\begin{equation}
  \dimlessP_\isomode(k) = \frac{k^3}{2\pi^2} \Sigma^{\isomode\isomode}(k)
  \quad \text{(no sum)} .
  \label{eq:dimlessP-sigma-def}
\end{equation}
If this typical amplitude is enhanced by a factor $E$,
the spatial variation of the long-wavelength
mode will be described by
\begin{equation}
  \delta \isomode(\vect{x}) \approx E \dimlessP_\isomode^{1/2}(\kL)
  \cos( \vectkL \cdot \vect{x} + \vartheta )
  \label{eq:modulating-mode}
\end{equation}
where $\vartheta$ is a phase
which will vary between realizations.

Contributions to the CMB anisotropy are generated
by $M$-patches located on the last scattering surface
at comoving distance $\xls$, \label{def:xls}
for which $\vect{x} = \xls \hat{\vect{n}}$.
(See Fig.~\ref{fig:LMpatch}.)
The unit vector
$\hat{\vect{n}}$ selects the line-of-sight from Earth.
If $\vectkL$ corresponds to a spatial scale
larger than $\xls$ then $\vectkL\cdot\vect{x} \lesssim 1$,
and in order that the linear gradient dominates we
should require $\vectkL\cdot\vect{x} \lesssim 10^{-1}$.
In principle there is no upper bound on the spatial
scale associated with $\kL$
because a reduction in the gradient can always be
compensated by adjusting the amplitude.
However, unless $\dimlessP_\isomode$ is rather red,
moving $\kL$ to larger scales will likely require
an increase in the exceptionality $E$.

In the Erickcek et al. scenario we should
prefer \emph{some} exceptionality
to justify our focus on a single wavenumber.
The alternative, that all modes have nearly equal
amplitude,
is
the scenario of
Schmidt \& Hui
and
Adhikari et al.~\cite{Schmidt:2012ky,Schmidt:2015xka,Adhikari:2015yya}.
But although some exceptionality is desirable
the probability of obtaining an
exceptional fluctuation decreases as
$E$ increases, and if we require $E$ to be very large the
scenario becomes unattractive.
We will generally assume that the best arrangement
is to set $E$
as small as possible,
but no lower than perhaps $\Or(10)$.
Notice that in doing so we are electing to trade a
$\sim 3\sigma$ discrepancy for
\emph{at least} a $\sim 10 \sigma$ fluctuation.
Therefore any realistic explanation of the
hemispherical asymmetry which deploys
this mechanism
will
likely require
some way to manufacture a suitable exceptional
amplitude without depending on Gaussian statistics.
Possible
examples include the proposals of Refs.~\cite{Liddle:2013czu,Mazumdar:2013yta}.

To parametrize the scale $\vectkL$ we write
\begin{equation}
	\vectkL = \frac{2\pi}{\xls} \alpha \hat{\vect{p}} ,	
	\label{eq:alpha-def}
\end{equation}
where $\hat{\vect{p}}$ is a unit vector and $\alpha < 1$
characterizes
the ratio of $\xls$ to the spatial scale associated with $\kL$.
After expanding in $\vectkL\cdot\vect{x}$,
Eq.~\eqref{eq:twopf-response}
gives an expression for the modulated two-point function,
\begin{equation}
	\langle
		\zeta(\vect{k}_1)
		\zeta(\vect{k}_2)
	\rangle_M
	\approx
	\langle
		\zeta(\vect{k}_1)
		\zeta(\vect{k}_2)
	\rangle_L
	\Big\{
		1
		- 2 C(k)
		+ 2A(k) \frac{\vect{x} \cdot \hat{\vect{p}}}{\xls}
		+ \cdots
	\Big\} ,
	\label{eq:twopf-modulation}
\end{equation}
where the quantities $A(k)$ and $C(k)$ are defined by
\begin{subequations}
\begin{align}
	\label{eq:A-def}
	A(k) & = \pi \alpha E \dimlessP_\isomode^{1/2}(\kL) \rho_\isomode(k) \sin \vartheta \\
	\label{eq:C-def}
	C(k) & = - \frac{1}{2}E \dimlessP_\isomode^{1/2}(\kL) \rho_\isomode(k) \cos \vartheta .
\end{align}	
\end{subequations}
This is of the required form~\eqref{eq:Pk-modulation} with
$\hat{\vect{n}} = \vect{x}/\xls$.
The scale-dependence of the modulation amplitude
$A(k)$ is inherited from
the scale-dependence of the response $\rho_\isomode(k)$.

\para{Suppression of low multipoles}%
Eq.~\eqref{eq:twopf-modulation}
shows that
in addition to the dipolar modulation
there is an overall shift in amplitude due to $C(k)$~\cite{Lyth:2014mga}.
Assuming the spatial dependence of the long-wavelength
mode is described by~\eqref{eq:modulating-mode},
this is related to $A(k)$ via
\begin{equation}
	C(k) = - \frac{A(k)}{2 \pi \alpha} \cot \vartheta .
	\label{eq:Ck-alpha-Ak}	
\end{equation}
For more general dependence
the coefficient of proportionality is altered, but
the scaling $C(k) \sim A(k)/\alpha$ remains.
In Refs.~\cite{Lyth:2013vha,Lyth:2014mga,McDonald:2014kia}
it was suggested
that $C(k)$ could be used to explain
a second CMB anomaly---the observed low
CMB quadrupole.
Schwarz et al. proposed that a
viable explanation of the hemispherical asymmetry should explain at
least one other anomaly, so this outcome would be desirable~\cite{Schwarz:2015cma}.
Unfortunately, Eq.~\eqref{eq:Ck-alpha-Ak}
will make $C(k)$ larger than required if $\alpha$ is small.
Assuming the reported BICEP2 measurement of $r \sim 0.2$
(now known to have been confused by dust),
Contaldi et al.
estimated that $C(k)$ could be
roughly of order $0.14$~\cite{Contaldi:2014zua}.
However, precise constraints do not seem to have been reported
in more general circumstances.
We will assume it should not be much larger than $\sim 0.1$,
and it should preferably not be negative.

This is an obstacle for construction of viable models.
If we assume the spatial dependence in~\eqref{eq:modulating-mode}
then
the amplitude $|C(k)|$ can be reduced by tuning $\vartheta$.
For more general spatial dependence
it requires tuning the
Taylor coefficient of order
$(\vectkL\cdot\vect{x})^0$
with respect to that of order
$(\vectkL\cdot\vect{x})^1$.

\para{Bi- and trispectrum amplitudes}%
If $\isomode$
dominates the bi- and trispectrum of $\zeta$
then it is possible to go further
and relate the amplitude of the
asymmetry $A(k)$
to the degree of correlation in,
respectively,
squeezed
and collapsed configurations
of the three- and four-point functions.
We will see in~\S\S\ref{sec:why-hard}--\ref{sec:working-model}
that this situation is realized in a large
class of successful scenarios.
To measure the amplitude of three-point
correlations we define the
reduced bispectrum $\fNL(k_1, k_2, k_3)$,
\begin{equation}
	\frac{6}{5} \fNL(k_1, k_2, k_3)
	=
	\frac{B(k_1, k_2, k_3)}{P(k_1)P(k_2) + P(k_1)P(k_3) + P(k_2)P(k_3)} ,	
	\label{eq:reduced-bispectrum}
\end{equation}
where $B(k_1, k_2, k_3)$ is the bispectrum for the
three-point function of $\zeta$,
\begin{equation}
	\langle
		\zeta(\vect{k}_1)
		\zeta(\vect{k}_2)
		\zeta(\vect{k}_3)
	\rangle_L
	=
	(2\pi)^3
	\delta(\vect{k}_1 + \vect{k}_2 + \vect{k}_3)
	B(k_1, k_2, k_3) .
	\label{eq:zeta-bispectrum}
\end{equation}
In the squeezed limit $k_3 \ll k_1, k_2$
this becomes approximately
\begin{equation}
  \frac{6}{5} \fNL(k, k, k_3)
  \approx
  \frac{B(k, k, k_3)}{2 P(k_3) P(k)}
  \quad
  \text{if $k \ll k_3$}
  . 
  \label{eq:reduced-bispectrum-squeezed}
\end{equation}
For four-point correlations we define
\begin{equation}
	\tauNL(\vect{k}_1, \vect{k}_2, \vect{k}_3, \vect{k}_4) \approx
	\frac{T(\vect{k}_1, \vect{k}_2, \vect{k}_3, \vect{k}_4)}{4 P(|\vect{k}_3 + \vect{k}_4|) P(k_2) P(k_4)}
	\quad
	\text{if $|\vect{k}_3 + \vect{k}_4| \ll k_2$, $k_4$}
	,
	\label{eq:tauNL-def}
\end{equation}
with $T(k_1, k_2, k_3, k_4)$ now the trispectrum defined by the
connected four-point
function of $\zeta$,
\begin{equation}
	\langle
		\zeta(\vect{k}_1)
		\zeta(\vect{k}_2)
		\zeta(\vect{k}_3)
		\zeta(\vect{k}_4)
	\rangle_L^\text{connected}
	=
	(2\pi)^3
	\delta(\vect{k}_1 + \vect{k}_2 + \vect{k}_3 + \vect{k}_4)
	T(\vect{k}_1, \vect{k}_2, \vect{k}_3, \vect{k}_4)
	.
	\label{eq:zeta-trispectrum}
\end{equation}

Taking our position on the long-wavelength mode
to be generic, so that $\sin \vartheta \sim \cos \vartheta \sim \Or(1)$,
Eq.~\eqref{eq:mixed-bispectrum-ope}
predicts that for observably squeezed configurations---%
including those which contribute to the CMB---%
we can write the reduced bispectrum in terms
of the amplitude $A(k)$,
\begin{equation}
	\frac{6}{5} \fNL(k,k,k_3) \approx
	\frac{A(k)}{2\pi \alpha E}
	R^{1/2}(k_3)
	\bigg(
	\frac{\dimlessP_\isomode(k_3)}{\dimlessP_\isomode(\kL)}
	\bigg)^{1/2}
	\frac{1}{\dimlessP^{1/2}(k_3)}
	\quad
	\text{if $\kL < k_3 \ll k$}
	,
	\label{eq:fNL-asymmetry}
\end{equation}
where
$\dimlessP(k)$ is the dimensionless version of the $\zeta$ power spectrum $P(k)$,
and
$R$ measures the contribution of $\isomode$
to the total power spectrum,
\begin{equation}
	R(k) \equiv
	\frac{\dimlessP_{\zeta_\isomode}(k)}{\dimlessP(k)}
	.
	\label{eq:R-def}
\end{equation}
Here $\zeta_\isomode$ is the linear contribution
of $\isomode$ to $\zeta$.
Despite its appearance~\eqref{eq:fNL-asymmetry} does not depend on
$\dimlessP^{1/2}(\kL)$ since
Eq.~\eqref{eq:A-def}
shows that $A(k)$
is proportional to it.
Similarly, Eq.~\eqref{eq:collapsed-trispectrum}
predicts that for collapsed configurations
\begin{equation}
	\tauNL(\vect{k}_1, \vect{k}_2, \vect{k}_3, \vect{k}_4)
	\approx
	\frac{A(k_2) A(k_4)}{4 \pi^2 \alpha^2 E^2}
	\frac{\dimlessP_\isomode(|\vect{k}_3 + \vect{k}_4|)}{\dimlessP_\isomode(\kL)}
	\frac{1}{\dimlessP(|\vect{k}_3 + \vect{k}_4|)}
	\quad
	\text{if $\kL < |\vect{k}_3 + \vect{k}_4| \ll k_2$, $k_4$}
	.
	\label{eq:tauNL-asymmetry}
\end{equation}
We caution that these relations hold \emph{only} if
a single field
$\isomode$ dominates the bi- and trispectra of $\zeta$.
Note that this does not restrict
Eqs.~\eqref{eq:fNL-asymmetry} and~\eqref{eq:tauNL-asymmetry}
to single-source models
in the sense of Refs.~\cite{Lyth:2013vha,Lyth:2014mga,Kobayashi:2015qma},
where the scale-dependent field $\isomode$
dominates all the correlation functions of $\zeta$,
because there is no requirement for $\isomode$
to dominate the two-point function.
This more general scenario has already been studied by
Kanno et al.~\cite{Kanno:2013ohv}.
They gave a formula equivalent to Eq.~\eqref{eq:fNL-asymmetry}
(although derived by a different method and not including explicit scale-dependence),
but we believe~\eqref{eq:tauNL-asymmetry}
is new.

Where
Eqs.~\eqref{eq:fNL-asymmetry} and~\eqref{eq:tauNL-asymmetry}
apply,
they
predict that the amplitudes
$\fNL$ and $\tauNL$ are enhanced
by inverse powers of $\alpha$
and the $\zeta$ power-spectrum
$\dimlessP$, but suppressed
by powers of the asymmetry $A$ and the
exceptionality $E$.

\subsection{Single-source models}
\label{sec:single-source}

Eqs.~\eqref{eq:twopf-modulation}, \eqref{eq:A-def}
and~\eqref{eq:C-def}
(and their obvious generalization to the case where
the modulation is sourced by perturbations in several fields)
explain how to calculate the asymmetry
produced by the Erickcek et al. scenario in an arbitrary local-like model.

Refs.~\cite{Lyth:2013vha,Lyth:2014mga,Kobayashi:2015qma} studied
biasing of the two-point function in
the special case of
single-source models
in the sense defined above.
These references reported
results equivalent to~\eqref{eq:A-def}
with $\rho_\isomode(k)$ replaced by
$12 \fNL(k,k,k) / 5$,
and $\dimlessP_\isomode(\kL)$
replaced by the
amplitude of $\zeta$ fluctuations,
$\dimlessP_\zeta(\kL)$.
Here $\fNL(k_1, k_2, k_3)$ is the reduced bispectrum
defined in Eq.~\eqref{eq:reduced-bispectrum}.

To make a connexion with the results of
Refs.~\cite{Lyth:2013vha,Lyth:2014mga,Kobayashi:2015qma}
we reformulate our analysis
in
terms of the response of the $\zeta$ two-point function
to a long-wavelength $\zeta$ fluctuation.
In a single-source model this leads to no loss of generality
since
we need not distinguish between
the field fluctuation which dominates $\zeta$,
and $\zeta$ itself.
In any model (single-source or not) the response
to a $\zeta$ fluctuation
can be obtained by projecting $\rho_\mu(k)$
along the field-space unit vector corresponding to the orientation
of $\zeta$.%
	\footnote{By `field space' we mean the
	space spanned by the field values and their momenta.}
To obtain this, note that $\zeta$ can always be expressed
as a composite of the field fluctuations defined on 
spatially flat hypersurfaces,
\begin{equation}
	\zeta(\vect{k})
	=
	N_\alpha \delta\phi^\alpha(\vect{k})
	+
	\frac{1}{2}
	N_{\alpha\beta}
	\int \frac{\d^3 q}{(2\pi)^3}
	\delta\phi^\alpha(\vect{q}) \delta\phi^\beta(\vect{k}-\vect{q})
	+
	\cdots ,
	\label{eq:gauge-transformation}
\end{equation}
where `$\cdots$' denotes terms of higher order in the field fluctuations
which have been omitted.
The fluctuations $\zeta$ and $\delta\phi^\alpha$ are to be evaluated
at the same time,
and the gauge transformation coefficients
$N_\alpha$, $N_{\alpha\beta}$
can be found in the literature~\cite{Guth:1982ec,Lyth:1984gv,Dias:2014msa}.
As above we caution that we are not using the slow-roll
approximation and therefore
the labels $\{ \alpha, \beta, \ldots \}$ run over both the
scalar fields and their momenta.

\para{Response to $\zeta$}%
At linear order $\zeta \approx N_\alpha \delta \phi^\alpha$.
Therefore we can associate the curvature perturbation with a fluctuation
oriented along the field-space unit vector
$\hat{n}_\alpha = N_\alpha / (N_\lambda N_\lambda)^{1/2}$.
In this expression and what follows,
summation over repeated field-space
indices is implied,
irrespective of their `up' or `down' position.
(We are working with a trivial field-space metric.)
It follows that the response to a long-wavelength $\zeta$
fluctuation can be computed by the projection
\begin{equation}
  \rho_\zeta(k)
  =
  \frac{\hat{n}_\alpha \rho_\alpha(k)}{(N_\lambda N_\lambda)^{1/2}}
  \approx
  \frac{N_\alpha [\Sigma^{-1}(k_3)]_{\alpha\mu} B^\mu(k, k, k_3)}
    {(N_\lambda N_\lambda) P(k)}
  \quad
  \text{if $k \ll k_3$}
  .
  \label{eq:zeta-response}
\end{equation}
The linear formula for $\zeta$ shows that
$P(k) = N_\alpha N_\beta \Sigma^{\alpha\beta}(k)$ up to corrections of
higher order in the field fluctuations.
Also, in this limit $B(k_1, k_2, k_3) \approx N_\mu B^\mu(k_1, k_2, k_3)$.
Taken together, these relations suggest that Eq.~\eqref{eq:zeta-response}
is
\emph{related}
to the squeezed limit
of the reduced bispectrum~\eqref{eq:reduced-bispectrum-squeezed}.
Nevertheless it is not exactly the same.
In Eq.~\eqref{eq:reduced-bispectrum-squeezed}
the combination $P(k_3) \approx N_\lambda N_\mu \Sigma^{\lambda\mu}(k_3)$
appears in the denominator,
whereas in Eq.~\eqref{eq:zeta-response}
only the contraction
$N_\lambda N_\lambda$ appears there directly;
the power spectrum factor appears in the numerator as a matrix
inverse interposed between $N_\alpha$ and $B^\alpha(k_1, k_2, k_3)$.

In the single-source, slow-roll limit these relations simplify.
The slow-roll approximation implies that $\zeta$ can be
written using only the fluctuation in the single relevant
field (without requiring the momentum fluctuation), so
matrix multiplication and inversion reduce
to ordinary multiplication and division.
In this limit the distinction between
Eqs.~\eqref{eq:zeta-response}
and~\eqref{eq:reduced-bispectrum-squeezed}
disappears, and
\begin{equation}
  \rho_\zeta(k) \approx \frac{B(k, k, k_3)}{P(k_3) P(k)}
  =
  \frac{12}{5} \fNL(k, k, k_3) . 
  \label{eq:zeta-response-single-source}
\end{equation}
In a single-source model the right-hand side
of~\eqref{eq:zeta-response-single-source}
can be computed and shown to be independent of
the soft mode $k_3$.
This observation follows from the formulae of Dias et al.~\cite{Dias:2013rla},
and was pointed out explicitly by Kenton \& Mulryne~\cite{Kenton:2015lxa}.
The final result agrees with the separate-universe
analyses
(obtained by entirely different methods)
presented in Refs.~\cite{Lyth:2013vha,Lyth:2014mga,Kobayashi:2015qma}.

Eq.~\eqref{eq:zeta-response-single-source}
and the property of independence from $k_3$
validate the statement made in Ref.~\cite{Lyth:2014mga},
that biasing of the $\zeta$ two-point function in single-source models
is controlled
by the reduced bispectrum on equilateral configurations.
In these models there is negligible difference between the reduced bispectrum
in equilateral and squeezed configurations of the same scale.

More generally, Eq.~\eqref{eq:response-function}
shows that the relevant configurations are squeezed rather than
equilateral, although
in local-like models
the response cannot depend on the squeezing
ratio $k_3/k_t$.
In an arbitrary model
it need \emph{not} happen that
$\fNL(k_1, k_2, k_3)$ becomes independent of $k_3$
in the limit $k_3 \ll k_1, k_2$,
which gives another demonstration that the response cannot equal
the reduced bispectrum in general.
Therefore to evaluate the amplitude of the asymmetry,
and its compatibility with Planck constraints on
the amplitude of three-point correlations in squeezed configurations,
will generally require a model-dependent analysis.
The same is true when checking the compatibility of a
Grishchuk--Zel'dovich effect
generated by an enhanced
mode of wavelength longer than the scale of the last-scattering surface.

\section{Why it is difficult to produce a suitable response}
\label{sec:why-hard}

Up to this point, it remains an open question whether it is
\emph{possible} to manufacture a response function
$\rho_\isomode(k)$ with suitable amplitude
and scale-dependence
without requiring an unacceptable
exceptionality $E$
or other undesirable features.
In this section we discuss, in general terms,
what properties appear required to yield
a scale-dependent bispectrum,
and explain why
the simplest scenarios have difficulty
in simultaneously producing an acceptable amplitude.

\subsection{Scale-dependence of the bispectrum}
\label{sec:scale-dependence}

Methods to compute the scale-dependence of
a bispectrum generated during inflation
have been discussed by several authors
\cite{Byrnes:2010ft,Byrnes:2012sc,Tzavara:2012qq}.
To describe the evolution of fluctuations on
scales outside the horizon we use the separate universe
picture to write an analogue of the
gauge transformation~\eqref{eq:gauge-transformation}
for the field fluctuations~\cite{Lyth:2005fi,Seery:2012vj}
\begin{equation}
	\delta \phi^\alpha(\vect{k})
	=
	\Gamma^\alpha_a \delta\phi^a(\vect{k})
	+
	\frac{1}{2} \Gamma^\alpha_{ab}
	\int \frac{\d^3 q}{(2\pi)^3}
	\delta\phi^a(\vect{q}) \delta\phi^b(\vect{k}-\vect{q})
	+
	\cdots ,
	\label{eq:separate-universe-fields}
\end{equation}
which again does not invoke the slow-roll approximation.
In this expression
the fluctuation on the left-hand side
is evaluated at the
time of interest but the fluctuations on the right-hand side
are evaluated at some earlier time.
To prevent our formulae becoming cluttered by a
proliferation of time labels
we indicate evaluation at this earlier time
by using the species labels $\{ a, b, \cdots \}$
instead of $\{ \alpha, \beta, \cdots \}$.

The mixed-index objects $\Gamma^\alpha_a$ and $\Gamma^\alpha_{ab}$
are derivatives of the background field configurations~\cite{Seery:2012vj},
\begin{subequations}
\begin{align}
	\Gamma^\alpha_a & = \frac{\partial \phi^\alpha}{\partial \phi^a} \\
	\Gamma^\alpha_{ab} & = \frac{\partial^2 \phi^\alpha}{\partial \phi^a \partial \phi^b}
	.
\end{align}	
\end{subequations}
The background solution $\phi^\alpha$ solves a system of
differential equations $\d \phi^\alpha / \d N = u^\alpha$.
Then, defining ${u^\alpha}_\beta = \partial_\beta u^\alpha$
and ${u^\alpha}_{\beta\gamma} = \partial_\gamma {u^\alpha}_\beta$
it is possible to write evolution equations for
$\Gamma^\alpha_a$ and $\Gamma^\alpha_{ab}$.
As functions of the time $N_+$ defined by
$\{ \alpha, \beta, \ldots \}$ they
obey~\cite{Yokoyama:2007dw,Yokoyama:2007uu,Yokoyama:2008by,Seery:2012vj,Anderson:2012em}
\begin{subequations}
\begin{align}
	\label{eq:gamma2-forward}
	\frac{\d}{\d N_+} \Gamma^\alpha_a
		& =
		{u^\alpha}_\beta \Gamma^\beta_a \\
	\label{eq:gamma3-forward}
	\frac{\d}{\d N_+} \Gamma^\alpha_{ab}
		& =
		{u^\alpha}_\beta \Gamma^\beta_{ab}
		+ {u^\alpha}_{\beta\gamma} \Gamma^\beta_a \Gamma^\gamma_b
 \end{align}	
\end{subequations}
whereas as functions of the time $N_-$ defined by
$\{ a, b, \cdots \}$ they obey
\begin{subequations}
\begin{align}
	\label{eq:gamma2-backward}
	\frac{\d}{\d N_-} \Gamma^\alpha_a
		& =
		- \Gamma^\alpha_b {u^b}_a \\
	\label{eq:gamma3-backward}
	\frac{\d}{\d N_-} \Gamma^\alpha_{ab}
		& =
		- \Gamma^\alpha_c {u^c}_{ab}
		- \Gamma^\alpha_{cb} {u^c}_a
		- \Gamma^\alpha_{ac} {u^c}_a
	.
\end{align}	
\end{subequations}

\para{Variation with scale}%
As we now explain,
these equations for the time-dependence of $\Gamma^\alpha_a$ and $\Gamma^\alpha_{ab}$ enable us
to determine the scale-dependence of correlation functions involving the
$\delta \phi^\alpha$.%
  \footnote{This method can be regarded as a refinement of the
  approach used in Refs.~\cite{Byrnes:2009pe,Byrnes:2010ft}.
  These references effectively constructed
  linear approximations
  for $\Gamma^\alpha_a$, $\Gamma^\alpha_{ab}$
  as a function of $N_-$, invoking the slow-roll approximation to
  control the expansion.
  These approximations were valid over only a small range of $k_t$.
  Eqs.~\eqref{eq:gamma2-backward} and~\eqref{eq:gamma3-backward}
  replace these approximations.
  They can be used to determine the $\Gamma$-matrices
  at any scale and, as we have explained, they do not require
  the slow-roll approximation.}
  
Eq.~\eqref{eq:separate-universe-fields}
permits us to write the two- and three-point functions
$\langle \delta\phi^\alpha \delta\phi^\beta \rangle$,
$\langle \delta\phi^\alpha \delta\phi^\beta \delta\phi^\gamma \rangle$
evaluated at the late time $N_+$ in terms of
$\langle \delta\phi^a \delta\phi^b \rangle$,
$\langle \delta\phi^a \delta\phi^b \delta\phi^c \rangle$
evaluated at the early time $N_-$,
giving
\begin{subequations}
\begin{equation}
	\langle
		\delta \phi^\alpha(\vect{k}_1)
		\delta \phi^\beta(\vect{k}_2)
	\rangle
	=
	\Gamma^\alpha_a \Gamma^\beta_b
	\langle
		\delta \phi^a(\vect{k}_1)
		\delta \phi^b(\vect{k}_2)
	\rangle
	\label{eq:fields-2pf}
\end{equation}
and
\begin{equation}
\begin{split}
	\langle
		\delta \phi^\alpha(\vect{k}_1)
		\delta \phi^\beta(\vect{k}_2)
		\delta \phi^\gamma(\vect{k}_3)
	\rangle
	=
	\mbox{}
	&
	\Gamma^\alpha_a \Gamma^\beta_b \Gamma^\gamma_c
	\langle
		\delta \phi^a(\vect{k}_1)
		\delta \phi^b(\vect{k}_2)
		\delta \phi^c(\vect{k}_3)
	\rangle
	\\
	&
	\mbox{} +
	\Gamma^\alpha_{mn}
	\Gamma^\beta_b
	\Gamma^\gamma_c
	\int \frac{\d^3 q}{(2\pi)^3}
	\langle
		\delta \phi^m(\vect{q})
		\delta \phi^b(\vect{k}_2)
	\rangle
	\langle
		\delta \phi^n(\vect{k_1}-\vect{q})
		\delta \phi^c(\vect{k}_3)
	\rangle
	\\
	&
	\mbox{} +
	\text{cyclic permutations}
	.
\end{split}
\label{eq:fields-3pf}
\end{equation}	
\end{subequations}

In what follows it is useful to make use of the scale
$k_t$ corresponding to the perimeter of the momentum $n$-gon
characterizing each correlation function.\label{def:kt}
For the two-point function $k_t = k_1 + k_2$ and for the
three-point function $k_t = k_1 + k_2 + k_3$, with obvious
generalizations to higher $n$-point functions.

To use Eqs.~\eqref{eq:fields-2pf} and~\eqref{eq:fields-3pf}
we set the time $N_-$
associated with $\{ a, b, \cdots \}$
to match the time at which the scale
$k_t$ exited the horizon in the sense $k_t / aH = 1$.
Now suppose we vary the momentum configuration,
first choosing to vary its scale $k_t$ while leaving
its shape (measured by the ratios
$k_i/k_t$) fixed.
This forces us to change the time $N_-$,
giving
two contributions to the variation
of the two- and three-point functions:
one arising from the change in evaluation time
for
$\Gamma^\alpha_a$ and
$\Gamma^\alpha_{ab}$,
which can be computed using Eqs.~\eqref{eq:gamma2-backward}--\eqref{eq:gamma3-backward};
and the other from the change
in the
two- and three-point functions
$\langle \delta \phi^a \delta \phi^b \rangle$,
$\langle \delta \phi^a \delta \phi^b \delta \phi^c \rangle$.

For configurations which are close to equilateral in the sense that
$k_i \sim k_t$ for each momentum $k_i$
the change from
$\langle \delta \phi^a \delta \phi^b \rangle$,
$\langle \delta \phi^a \delta \phi^b \delta \phi^c \rangle$
can be obtained using known results for the $n$-point
functions~\cite{Seery:2005gb}.
The dominant scaling is $1/k_t^3$ for the two-point function
and $1/k_t^6$ for the three-point function,
following directly from their engineering dimension but
cancelling out in dimensionless combinations such as the
reduced bispectrum.
The remaining scaling comes mostly from the time dependence of
the Hubble parameter $H$ or the field momenta $\d \phi^\alpha / \d N$
evaluated at $N_-$.
The contribution to~\eqref{eq:fields-3pf}
from $\langle \delta \phi^a \delta \phi^b \delta \phi^c \rangle$
is known to be negligible unless the fields have nontrivial derivative
interactions~\cite{Vernizzi:2006ve,Lyth:2005qj},
and
we discard it in the following discussion.

With these assumptions we can estimate the scaling with $k_t$
by temporarily invoking the slow-roll approximation to simplify
our expressions.
However, our conclusions will not depend on these simplifications.
When we discuss a concrete model
in~{\S}\ref{sec:working-model} we will employ a numerical method
which does not use the slow-roll approximation.

The slow-roll approximation
makes the variation of $H$ negligible when the scale $k_t$ was leaving the
horizon (even if slow-roll was subsequently violated),
so the only contribution which needs to be kept is that generated
by the $\Gamma$-matrices.
Also, we do not need to retain
fluctuations in the scalar field momenta
and therefore
we can restrict the indices on
$\Gamma^\alpha_a$, $\Gamma^\alpha_{ab}$,
${u^\alpha}_\beta$
and ${u^\alpha}_{\beta\gamma}$
to the field fluctuations only.
With this understanding we have
\begin{subequations}
\begin{align}
	{u^\alpha}_\beta
	& =
	- \frac{V_{\alpha\beta}}{3H^2}
	+ \Big(1 - \frac{\epsilon}{3} \Big)
		\frac{\dot{\phi}^\alpha \dot{\phi}^\beta}{H^2 \Mp^2}
	+ \frac{2}{3} \frac{\dot{\phi}^{(\alpha} \ddot{\phi}^{\beta)}}{H^3 \Mp^2}
	\\
	{u^\alpha}_{\beta\gamma}
	& =
	- \frac{V_{\alpha\beta\gamma}}{3H^2}
	+ \frac{\dot{\phi}^\alpha}{H \Mp} {u^\beta}_\gamma
	+ \frac{\dot{\phi}^\beta}{H \Mp} {u^\alpha}_\gamma
	+ \frac{\dot{\phi}^\gamma}{H \Mp} {u^\alpha}_\beta
	- \frac{\dot{\phi}^\alpha \dot{\phi}^\beta \dot{\phi}^\gamma}{H^3 \Mp^3} 
\end{align}
\end{subequations}
(in which the placement of indices should be regarded as
immaterial, owing to our use of a trivial field-space metric).
The quantity $\epsilon$ is the usual slow-roll parameter
$\epsilon \equiv - \dot{H}/H^2$.
If all fields are slowly rolling then $\dot{\phi}^\alpha / H \Mp \ll 1$,
making ${u^\alpha}_\beta$ principally sensitive to $V_{\alpha\beta}$
and ${u^\alpha}_{\beta\gamma}$ principally sensitive to $V_{\alpha\beta\gamma}$.

\para{Obtaining power-law scaling}%
In order to be concrete we restrict attention to
the scenarios
considered at the end of~\S\ref{sec:perturbed-2pf}
in which only a single field $\isomode$
dominates the $\zeta$ bispectrum.
These allow the simplest
possible statement.
However we expect that our qualitative conclusions continue to apply
in more general models.

First suppose that the second derivative of the potential
in the direction $\isomode$ is large compared to $H$,
while the third derivative is small.
Then ${u^\isomode}_\isomode \approx - \eta_\isomode$ while ${u^\isomode}_{\isomode\isomode} \approx 0$,
where $\eta_\isomode = V_{\isomode\isomode} / 3H^2$.
It follows that
\begin{subequations}
\begin{align}
	\label{eq:large-eta-2}
	\frac{\d}{\d \ln k_t} \Gamma^\isomode_\isomode & \approx \eta_\isomode \Gamma^\isomode_\isomode \\
	\label{eq:large-eta-3}
	\frac{\d}{\d \ln k_t} \Gamma^\isomode_{\isomode\isomode} & \approx 2 \eta_\isomode \Gamma^\isomode_{\isomode\isomode} .
\end{align}	
\end{subequations}
If $\eta_\isomode$ is approximately constant while
some range of wavenumbers
are leaving the horizon
then we conclude that, over this range, $\Gamma^\isomode_\isomode \sim k_t^{\eta_\isomode}$
and $\Gamma^\isomode_{\isomode\isomode} \sim k_t^{2\eta_\isomode}$.
This is the simplest mechanism
by which one can generate significant scale dependence.
The price we must pay to achieve any desired power-law scaling is a
tuning of the mass, together with the necessity to keep this mass
nearly constant over the desired range of scales.
To obtain a red-tilted power law we require $\eta_\isomode < 0$,
so during this epoch the field $\isomode$ can be regarded as
departing from a quadratic hilltop.
The principal disadvantage of this mechanism is that it affects both
$\Gamma^\isomode_\isomode$
and $\Gamma^\isomode_{\isomode\isomode}$
and therefore both the two- and three-point functions
will exhibit scale dependence.
This makes model-building more complex because the $\isomode$
contribution to the $\zeta$ two-point function must be kept
sufficiently small that the resulting spectral index is
acceptable.

Alternatively, in some scenarios it may be possible
for ${u^\isomode}_{\isomode\isomode}$ to be large
while ${u^\isomode}_\isomode$ remains small.
It is normally difficult to maintain this situation
over many $e$-folds, because large contributions
to ${u^\alpha}_{\beta\gamma}$
typically source large contributions to
${u^\alpha}_\beta$.
However, supposing it can be realized
this scenario
will generate scaling which satisfies
\begin{subequations}
\begin{align}
	\frac{\d}{\d \ln k_t} \Gamma^\isomode_\isomode & \approx 0 \\
	\label{eq:xi-equation}
	\frac{\d}{\d \ln k_t} \Gamma^\isomode_{\isomode\isomode} & \approx \frac{\xi_\isomode}{\Mp} \Gamma^\isomode_\isomode ,
\end{align}	
\end{subequations}
where we have set $\xi_\isomode = \Mp V_{\isomode\isomode\isomode}/ 3H^2$.
Compared to the large-$\eta_\isomode$
case it is less simple to obtain a pure power-law,
although by choosing $\xi_\isomode$
appropriately it is still possible to generate
scale dependence.
The advantage of this scenario is that the two-point function
does not acquire significant scale-dependence.

If we were to abandon the slow-roll approximation
then similar conclusions would apply if, respectively,
${u^\isomode}_\isomode$
or ${u^\isomode}_{\isomode\isomode}$
are larger than the other components of ${u^\alpha}_\beta$
and ${u^\alpha}_{\beta\gamma}$.
In the case of multiple fields a similar discussion will
apply to each field individually
unless the $u$-matrices couple the scale-dependent
species.

In either case, once
a set of scaling behaviours have been generated
by Eqs.~\eqref{eq:gamma2-backward} and~\eqref{eq:gamma3-backward},
the subsequent evolution cannot generate new ones.
Eqs.~\eqref{eq:gamma2-forward} and~\eqref{eq:gamma3-forward}
\emph{do} allow the evolution to vary the scaling observed in any particular
correlation function
by linearly
mixing the available scalings with new amplitudes---%
but it is only these amplitudes which depend on the superhorizon
epoch and not the scale-dependence itself.
This conclusion is quite general and does not depend on the
slow-roll approximation.

\para{Variation with shape}%
We now return to the alternative possibility
of variations in the momentum configuration which leave $k_t$
fixed but vary the side ratios $k_i/k_t$.
We describe this as a change of shape.
It is relevant only for the three- and higher $n$-point functions.

A variation in shape does not alter the evaluation time $N_-$
for the coefficients $\Gamma^\alpha_a$, $\Gamma^\alpha_{ab}$
in Eqs.~\eqref{eq:fields-3pf}.
It will change only the three-point function
$\langle \delta\phi^a(\vect{k}_1) \delta\phi^b(\vect{k}_2) \delta\phi^c(\vect{k}_3) \rangle$
and a subset of the two-point functions
appearing in the cyclically-permuted quadratic terms.
The effect on the two-point functions is to change their
evaluation time relative to the horizon-crossing time for their wavenumber,
and Eq.~\eqref{eq:fields-2pf}
shows that this can be expressed in terms of the 2-component $\Gamma$ coefficient
evaluated between suitable times.
The effect on the three-point function is more difficult to extract
but has recently been calculated by Kenton \& Mulryne~\cite{Kenton:2015lxa}.
It can also be expressed purely in terms of the 2-component $\Gamma$ coefficient.

The conclusion is that models which generate a significant scale-dependence
through a large $\eta_\isomode$ will
almost always exhibit strong scaling
as a function of the squeezing $k_i/k_t$.
Together with the scaling with $k_t$,
this will have implications for the degree to which we can interpret
recent Planck measurements of
$\fNLlocal$
(or the amplitude of other templates)
as measurements of the correlation amplitude in squeezed configurations.
Conversely, models which generate scale-dependence
through a large $\xi_\isomode$ while keeping $\eta_\isomode$
small
will exhibit much smaller scaling as a function of squeezing
because the 3-component $\Gamma$ coefficient is not needed to
describe scaling with shape at fixed $k_t$.

\subsection{Models with constant $\eta_\isomode$ are not viable}
\label{sec:no-const-eta}

In this paper we focus on
the simpler mechanism which generates
significant $k_t$-scaling
using a large second derivative, as in
Eqs.~\eqref{eq:large-eta-2} and~\eqref{eq:large-eta-3}.
On the basis of what has been said above, we should expect
this $k_t$-dependence to be accompanied by significant squeezing-dependence.

\para{Scaling of $\fNL$}%
How large an $|\eta_\isomode|$ is required?
We focus on a simple model in which inflation is driven
by a field $\phi$ which acquires a nearly
scale-invariant spectrum
while a second field
$\isomode$ acquires a scale-dependent
fluctuation using Eqs.~\eqref{eq:fields-2pf} and~\eqref{eq:fields-3pf}.
The simplest arrangement occurs if the $\isomode$
fluctuation
contributes significantly to the $\zeta$ three-point function
but not its two-point function,
because then the two-point function can be insulated
from any large scale-dependent contribution carried
by $\delta\isomode$.

First consider variations of scale $k_t$.
In the circumstances described above,
$\langle \zeta \zeta \zeta \rangle$
will scale like
$\langle \delta \isomode \delta \isomode \delta \isomode \rangle
\sim k_t^{-6} (\Gamma^\isomode_\isomode)^2 (\Gamma^\isomode_{\isomode\isomode})^2 \sim k_t^{-6+4\eta_\isomode}$
whereas
the scaling of
$\langle \zeta \zeta \rangle$
will be independent, like
$k_t^{-3+(n_s-1)}$.
Therefore to lowest order in slow-roll parameters
\begin{equation}
  	\fNL \sim k_t ^{4\eta_\isomode - 2(n_s-1)}
  	\quad
  	\text{(variation of $k_t$, shape fixed)}.
  	\label{eq:fNL-kt-slow-roll}
\end{equation}
Our approximations make this prediction independent of shape,
although in principle
shape-dependent contributions are present
through the correlation functions
$\langle \delta\phi^a \delta\phi^b \rangle$,
$\langle \delta\phi^a \delta\phi^b \delta\phi^c \rangle$
which we have neglected.
This property was noticed in Ref.~\cite{Byrnes:2010ft},
and
we briefly reconsider it in~\S\ref{sec:bsp-shape} below.

For measurements of the response function we
are instead interested in squeezed isosceles configurations
as described in~\S\ref{sec:ope}.
On squeezed configurations the results of
Kenton \& Mulryne
together with Eq.~\eqref{eq:reduced-bispectrum-squeezed}
suggest
$\langle \zeta \zeta \zeta \rangle
\sim
\langle \delta\isomode \delta\isomode \delta\isomode \rangle
\sim k_3^{-3} (\Gamma^\isomode_\isomode)^2 \sim k_3^{-3+2\eta_\isomode}$
(where now the time $N_-$ appearing in the $\Gamma$-matrix is to be
interpreted as the horizon exit time for $k_3$~\cite{Kenton:2015lxa})
while the scaling of the $\zeta$ two-point function is still
$\langle \zeta \zeta \rangle \sim k_3^{-3+(n_s-1)}$.
Therefore
\begin{equation}
	\fNL \sim
	\bigg(
		\frac{k_3}{k_t}
	\bigg)^{2\eta_\isomode - (n_s-1)}
	\quad
	\text{(variation of shape, $k_t$ fixed)}
	.
	\label{eq:fNL-squeezing-slow-roll}
\end{equation}
Putting Eqs.~\eqref{eq:fNL-kt-slow-roll}
and~\eqref{eq:fNL-squeezing-slow-roll}
together
enables us to estimate the scaling on a sequence
of isosceles triangles which fix $k_3$
but neither
$k_t$ nor $k_3/k_t$,
\begin{equation}
	\fNL(k,k,k_3) \sim
	k^{4\eta_\isomode-2(n_s-1)}
	\bigg( \frac{k_3}{k} \bigg)^{2\eta_\isomode-(n_s-1)}
	\sim
	k^{2\eta_\isomode-(n_s-1)}
	\quad
	\text{if $k \gg k_3$}
	\label{eq:fNL-isosceles-slow-roll}
	.
\end{equation}

Eq.~\eqref{eq:fNL-asymmetry}
shows that, under our assumptions,
the asymmetry amplitude $A(k)$ will scale like
$\fNL(k,k,k_3)$ at fixed $k_3$.
Therefore we can estimate the $\eta_\isomode$ required to generate a fixed
power law.
The scalar spectral index $n_s$ is known from
observation to be of order $n_s \approx 0.97$~\cite{Ade:2015lrj},
and
to obtain a power-law for $\fNL(k,k,k_3)$ which is a little less
steep than $k^{-0.5}$ we
will take $\eta_\isomode \approx -0.2$.

\para{Ridge models}%
This choice for $\eta_\isomode$
is attractive, because its relative
largeness provides a means to synthesize
three-point correlations with sufficient amplitude to modulate the
$\zeta$ two-point function.
Elliston et al. showed
that an inflationary trajectory
initially parallel to but slightly displaced from
a quadratic `ridge' in the inflationary
potential
generates an enhanced bispectrum
when the trajectory turns, eventually becoming nearly perpendicular
to its original direction of travel~\cite{Elliston:2011dr}.
The amplitude of the bispectrum at the point of maximum enhancement
is proportional to the $\eta$ parameter characterizing the ridge.
Moreover, the ridge will naturally generate a \emph{negative}
$\eta$ and therefore a red-tilted power-law.

\begin{figure}
	\begin{center}
		\includegraphics[scale=0.85]{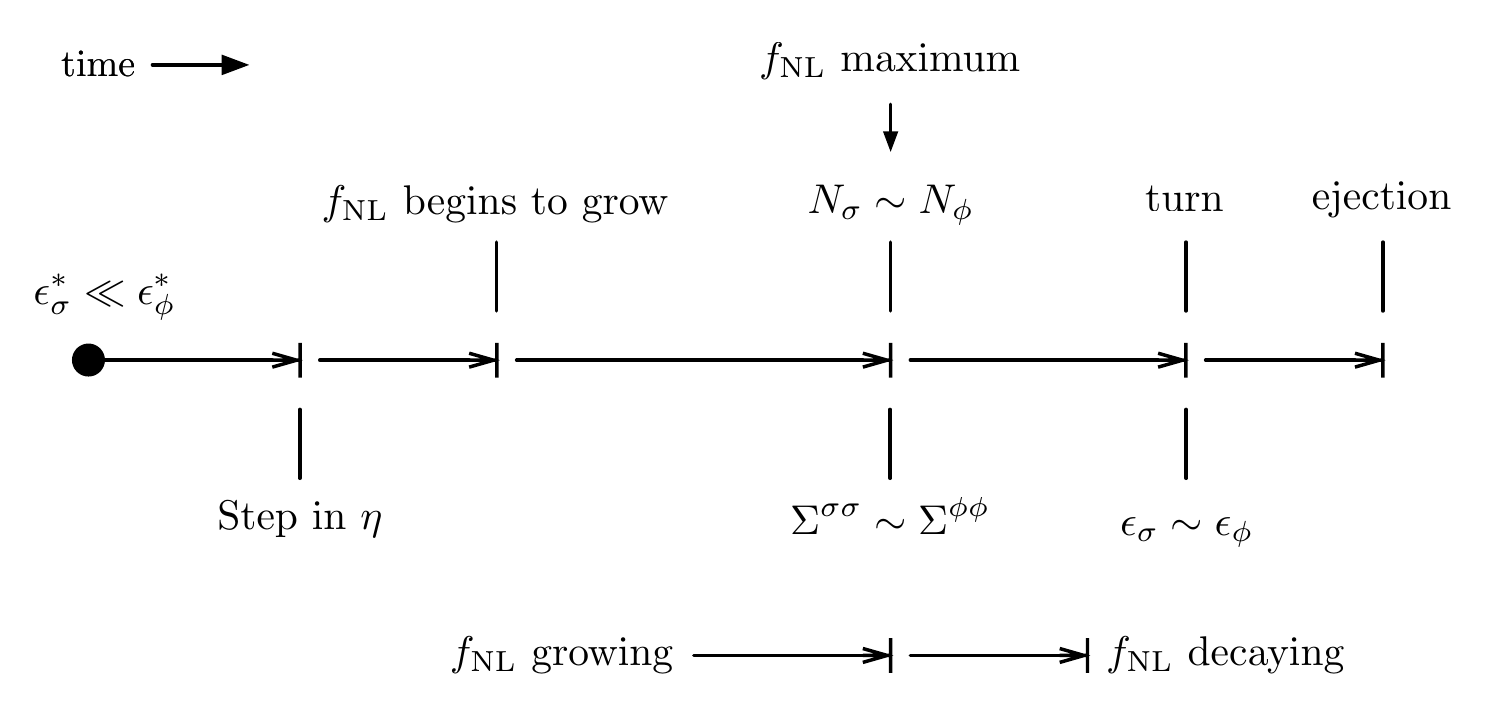}
	\end{center}
	\caption{Evolution of $\fNL$ generated by dispersion from
	a ridge with constant $\eta$.
	\label{fig:timeline}}
\end{figure}

The evolution of the bispectrum amplitude during this process is
depicted in Fig.~\ref{fig:timeline}.
The direction parallel to the ridge is the inflaton direction $\phi$
and the perpendicular direction is $\isomode$.
We measure the relative kinetic energies by writing
$\delta = \dot{\isomode}/\dot{\phi} = \sqrt{\epsilon_\isomode / \epsilon_\phi}$,
where
the $\epsilon$-parameters are defined by
\begin{subequations}
\begin{align}
	\epsilon_\phi & \equiv \frac{\Mp^2}{2} \left( \frac{V_\phi}{V} \right)^2 \\
	\epsilon_\isomode & \equiv \frac{\Mp^2}{2} \left( \frac{V_\isomode}{V} \right)^2 .
\end{align}
\end{subequations}
Labelling evaluation at the initial time by `$\ast$',
the initial conditions are chosen
to set up a significant kinetic-energy imbalance
$\epsilon_\isomode^\ast \ll \epsilon_\phi^\ast$,
and therefore $\delta_\ast \ll 1$.
This kinetic energy imbalance implies that
$\phi$ dominates $\zeta$.
For example, one can verify that the gauge transformation
coefficients $N_\alpha$ defined in~\eqref{eq:gauge-transformation}
satisfy $N_\isomode^\ast \sim \sqrt{\delta_\ast} N_\phi^\ast$,
and therefore $N_\isomode^\ast \ll N_\phi^\ast$.

Eventually the trajectory will begin to depart from the ridge
and the kinetic energy imbalance will begin to equalize.
During this process
the contribution of $\delta\isomode$ to $\zeta$ at a fixed
scale becomes both more significant and more nonlinear,
while the $\delta\phi$ contribution is largely unchanged.
This causes the $\zeta$ bispectrum to grow.
The peak amplitude is inversely proportional
to the original imbalance,
of order
$|\eta_\isomode / \delta_\ast|$.
It is typically reached at the
point where both $\delta\phi$ and $\delta\isomode$
make comparable contributions to the $\zeta$
two-point function. This occurs while $\delta < 1$,
and therefore before the point of equal kinetic energy
which we describe as `the turn' in Fig.~\ref{fig:timeline}.
By the time the trajectory turns the bispectrum amplitude
is already decaying.
After the turn
the evolution depends
on the precise form of the potential at large distances from the ridge,
labelled `ejection' in Fig.~\ref{fig:timeline}.

This process is a close analogue of the curvaton mechanism,
which relies on equalizing a large initial \emph{energy density} imbalance
between $\phi$ and $\isomode$
rather than a kinetic energy imbalance.
The equalization phase typically happens after inflation
because the $\isomode$ potential need not be of inflationary type.
Here also
the final amplitude is typically inversely proportional to the
initial imbalance, but in a curvaton model the $\delta \isomode$
fluctuation would normally come to dominate both the
$\zeta$ two- and three-point functions.
If the $\delta \isomode$ spectrum is too far from scale invariance
this will lead to an unacceptable spectral index.
Therefore, in either case,
we have to contrive an exit
which occurs when $\delta\isomode$
contributes to the $\zeta$ three-point function
but does not dominate the two-point function.
We focus on the ridge case
on the assumption that it will be less easy to realize
a working curvaton scenario.

\para{Evolution of $\isomode$}%
We now consider the evolution of $\isomode$ while the bispectrum
is beginning to grow.
Recall that, in this section, we are still temporarily imposing
the slow-roll approximation in order to simplify our formulae.

While the evolution of $\isomode$ is described by a constant
$\eta$-parameter, it will grow like
\begin{equation}
  \isomode \approx \isomode_\ast \e{-\eta_\isomode N} .  
\end{equation}
We have assumed that the ridge lies at $\isomode = 0$
and that the initial conditions displaced $\isomode$ to slightly
positive values, so that
$\isomode$ will roll in this
direction
at later times.
During this era the reduced bispectrum amplitude
on the equilateral configuration
which left the horizon at the initial time
will grow approximately
like~\cite{Elliston:2011dr,Elliston:2011et}
\begin{equation}
  \frac{6}{5} \fNL \approx - \frac{\eta_\isomode}{\delta_\ast} R^3 ,  
  \label{eq:ridge-fNL-grow}
\end{equation}
where $R$
is the quantity defined in~\eqref{eq:R-def}.
It measures the relative contribution
of the $\isomode$ and $\phi$ fluctuations
to the $\zeta$ two-point function.
Eq.~\eqref{eq:ridge-fNL-grow} can be interpreted as a constraint on
the initial kinetic energy imbalance given a value for $\eta_\isomode$
and a desired amplitude $\fNL$.
But we cannot make the initial imbalance too extreme
without positioning $\isomode_\ast$ very close to the top of the ridge
where its evolution is dominated by quantum diffusion rather
than classical evolution.
To stay outside the diffusion regime we require
the classical motion in a single e-fold
$\sim \d \isomode / \d N$
to dominate the quantum motion
$\delta \isomode / \delta N \sim H/2\pi$.
That gives
$2 \Mp^2 \epsilon_\isomode^\ast \gtrsim (H_\ast/2\pi)^2$,
or equivalently
\begin{equation}
    \delta_\ast^2 = \frac{\epsilon_\isomode^\ast}{\epsilon_\phi^\ast}
    \gtrsim \frac{1}{2} \frac{H_\ast^2}{4\pi^2} \frac{1}{\Mp^2 \epsilon_\phi^\ast}
    \approx
    \frac{\dimlessP}{2} .
\end{equation}
In the final step we have used the approximation that the
$\phi$ contribution to $\dimlessP$ barely evolves while
$\fNL$ is growing.
If necessary it is possible to make a more precise estimate,
but the outcome is hardly altered.
Therefore we conclude
\begin{equation}
  \frac{6}{5} |\fNL| \ll
  \bigg|
  \frac{\sqrt{2} \eta_\isomode R^3}{\dimlessP^{1/2}}
  \bigg| .
  \label{eq:fNL-constraint}
\end{equation}
We have written `$\ll$' because to obtain an
acceptable outcome it
is necessary to take $\delta_\ast$ substantially
larger than the lower limit $\sim \dimlessP^{1/2}$.
This is because
initial conditions which are too close to the
diffusion regime
would lead to very large fluctuations on the
largest observable scales.
For $\eta_\isomode \sim 0.2$
and $R \sim 0.1$
Eq.~\eqref{eq:fNL-constraint} gives $|\fNL| \ll 5$,
which implies that it will hardly be possible
to achieve even $\fNL \sim 1$.
If we require more stringent bounds on $R$ then the
situation is worse.
The conclusion is that, if one seeks to synthesize
a scale-dependent bispectrum
during inflation using the ridge mechanism with constant
$\eta_\isomode$,
and insists that the $\isomode$ two-point function
remains subdominant in the $\zeta$
power spectrum,
it will not be possible to start with $\delta_\ast$
sufficiently small to
to yield a large amplitude.

\para{Alternative strategies}%
This argument invokes relatively
strong
assumptions,
such as a constant $\eta_\isomode$,
and does not preclude the possibility that a working model can
be found.
Instead it should be regarded as a guide,
suggesting that in a large-$\eta_\isomode$ model \emph{either}
the $\isomode$ contribution to the $\zeta$ two-point function
cannot remain always subdominant
\emph{or}
that the $\isomode$ potential cannot be featureless.
These suggestions highlight generic difficulties.
First,
the $\isomode$ power spectrum will typically be strongly
scale-dependent if $\eta_\isomode$ is large enough to generate
a scale-dependent bispectrum.
Second, features in the $\isomode$ potential
will typically imply fine-tuning.
In the remainder of~\S\S\ref{sec:why-hard}--\ref{sec:working-model}
we develop these difficulties in more detail.
A related discussion has been given
by Kanno et al.~\cite{Kanno:2013ohv}.

Specifically, attempts to construct a working model
using a large value for $\eta_\isomode$ normally encounter at least
one of the following difficulties.

\begin{itemize}
	\item If the $\isomode$ potential
	is featureless and we terminate the inflationary
	phase before the
	scale-dependent $\delta\isomode$
	two-point function
	contributes significantly to the $\zeta$ two-point function,
	in order to protect near scale-invariance,
	then we will not normally generate a sufficiently
	large bispectrum to produce the desired response.
	
	\item To allow the development of large amplitudes
	one can add features to the $\isomode$ power spectrum.
	This will be the approach taken in~\S\ref{sec:working-model}
	below.
	Models of this kind may be successful but typically require
	several tunings,
	including at least the details of the feature
	and the exit from inflation.
	The exit point must be fine-tuned so that the bispectrum amplitude
	is sufficiently large but $R$ is still sufficiently small.
	Even if this can be done
	there may be issues with the amplitude of the
	trispectrum, to be described below,
	and the required initial conditions may still be
	uncomfortably close to the diffusion regime.
	
	\item Alternatively one can abandon the idea
	of keeping $R$ always small,
	allowing~\eqref{eq:ridge-fNL-grow} to yield
	a larger amplitude.
	This could be done by retaining
	the ridge mechanism but allowing $\fNL$
	to pass through the point of maximum amplitude,
	or by extending the model to include a subsequent
	curvaton era.
	
	In the ridge case one must still tune the exit from inflation
	to occur at the correct amplitude,
	while also finding some mechanism to suppress
	$R$ later. We expect this requires further tuning.
	In the curvaton case there is no need to tune
	the inflationary exit,
	but it seems even more difficult to suppress
	$R$ once the curvaton has come to dominate the
	energy density.
\end{itemize}

A different strategy would be to abandon the large-$\eta_\isomode$
method for obtaining scaling in the bispectrum,
instead attempting to find
a suitable $\xi_\isomode$ while keeping $\eta_\isomode$ very small.
This makes it easier
to keep the $\delta\isomode$ two-point function
nearly scale-invariant,
but one would have to contrive a suitable
$\xi_\isomode$ evolution which gave a power-law
response \emph{and}
did not induce a large $\eta_\isomode$ at any
point during the evolution.
Although we have not proved that this cannot be done,
it is far from easy to do so.
  
\para{Trispectrum constraints}%
The simplest possibility is to keep $R < 1$
and tune the time of exit from inflation.
However, as we now explain, this
typically leads to a large amplitude
in the $\tauNL$ mode of the trispectrum.

In a model which develops a large bispectrum amplitude through
superhorizon evolution there will typically be two
contributions to the trispectrum:
a $\gNL$-mode, which is small for models
without large cubic self-interactions~\cite{Seery:2006vu,Seery:2008ax,Seery:2006js,Byrnes:2006vq},
and a $\tauNL$-mode
(discussed in~\S\ref{sec:perturbed-2pf} above)
which obeys the Suyama--Yamaguchi relation~\cite{Suyama:2007bg},
\begin{equation}
	\tauNL \geq
	\Big(
		\frac{6}{5} \fNL
	\Big)^2 .
\end{equation}
In a model with significant scaling the precise momentum dependence
associated with these trispectrum contributions
will be modified in comparison with the standard templates, but
we expect their basic character to be preserved.

Equality occurs for single-source models.
In multiple-source models
the amplitude of the $\tauNL$ shape
may be enhanced.
Under the same assumptions which led to
the estimates~\eqref{eq:fNL-asymmetry}
and~\eqref{eq:tauNL-asymmetry}
for $\fNL$ and $\tauNL$
in terms of $A(k)$
we can infer a relationship between
$\fNL$ evaluated on squeezed configurations
and $\tauNL$ evaluated on collapsed configurations
of a similar scale,
\begin{equation}
	\label{eq:tauNL-estimate}
	\tauNL
	\simeq
	\frac{1}{R^2}
	\Big(
		\frac{6}{5} \fNL
	\Big)^2
	.
\end{equation}
Therefore the trispectrum amplitude is enhanced
by a factor $1/R^2$.
This can be understood if
we interpret~\eqref{eq:fNL-asymmetry}
to mean that the price paid to obtain
a strongly scale-dependent bispectrum is suppression
of $\fNL$ below its natural value by the factor $R^{1/2} < 1$.
However, $\tauNL$ is not suppressed in the same way
and is therefore substantially larger than the
lower bound provided by the Suyama--Yamaguchi relation~\cite{Byrnes:2015asa}.

If we choose $R \ll 1$ to protect the near scale-invariance
of the power spectrum then the enhancement can be considerable.
The same estimate~\eqref{eq:tauNL-estimate}
can be obtained using slow-roll results for the amplitude
of $\tauNL$ on tetrahedral configurations.%
	\footnote{A `tetrahedral' configuration is
	a special case of an equilateral configuration
	for which $k_i = |\vect{k}_i - \vect{k}_j| = k$
	for some $k$ and all $i \neq j$. This is the appropriate
	trispectrum generalization of an equilateral
	bispectrum configuration, in the sense that it involves
	only a single scale. Therefore large effects
	due to a hierarchy of scales cannot enter~\cite{Dias:2012qy}.}
These additionally enable us to estimate
the scale-dependence of $\tauNL$ on such configurations,
\begin{subequations}
\begin{align}
	\label{eq:ntauNL-estimate}
	\ntauNL
	& \simeq
		6 (\eta_\isomode - \eta_\phi )
		\simeq
		\frac{3}{2} \nfNL ,
\end{align}	
\end{subequations}
where $\nfNL$, $\ntauNL$ refer to the spectral index
of the reduced bispectrum
and the amplitude of the $\tauNL$-shape with variations of $k_t$.

Models which achieve strong scaling through a large
$\xi_\isomode = \Mp^3 V_{\isomode\isomode\isomode}/V$
are instead likely to generate large $\gNL$.
Ref.~\cite{Byrnes:2006vq} studied a concrete example
based on a self-interacting curvaton, showing
that it brought the model into conflict with
observational constraints
both for $\gNL$ and the quadrupolar asymmetry of the power spectrum.
To estimate how large a $\xi_\isomode$ might be
required to generate a relevant scale-dependence
we use the slow-roll results given in Byrnes et al.~\cite{Byrnes:2010ft},
which yield roughly $|\xi_\isomode| \gtrsim 100$.
(Notice that this does not imply a breakdown of
the slow-roll conditions for $\isomode$,
although as explained above
we would expect a large $\xi_\isomode$
to generate a large $\eta_\isomode$
which in turn might conflict
with scale-invariance of the
$\zeta$ power spectrum.)
In~\S\ref{sec:conclusions} we comment on further challenges arising
in large $\xi_\isomode$ models.

\para{Non-inflationary scenarios}%
Finally one might consider whether
scenarios exist which do not rely
on a large-amplitude bispectrum to couple long- and short-scale
perturbations.
Examples include: a domain wall~\cite{Jazayeri:2014nya};
`thawing' cosmic strings~\cite{Ringeval:2015ywa};
parity violating fluctuations in the initial
state~\cite{Ashoorioon:2015pia};
a modulation of the long wavelength mode, coupling
to isocurvature perturbations which persist
until at least decoupling~\cite{Assadullahi:2014pya};
or spatially-modulated dissipation of the
inflaton during inflation \cite{D'Amico:2013iaa}.

Some of these models are attractive because they naturally explain
the scale dependence of the asymmetry, whereas in the inflationary case it
requires a model-building choice to realize a large $\eta_\isomode$
or $\xi_\isomode$.
However, this does not imply that the resulting scale-dependence
will match observation.
For example, the simplest realization of
the domain wall model predicts
$A(k)\propto1/k$,
which is too steep to match the data.
But more
importantly (to the best of our knowledge) none of
these models have been shown explicitly to match
all the observational constraints,
including those of the bi- and trispectrum,
the low-$\ell$ multipoles of the CMB,
\emph{and} a quadrupolar modulation of
the power spectrum. In many cases the shape and amplitude
of the non-Gaussianity has not been computed
or compared to observation---although this criticism applies equally
to the inflationary case, which we will take up in~\S\ref{sec:working-model}
below.

\section{A working but contrived model}
\label{sec:working-model}

In this section we illustrate the difficulties highlighted
in~{\S}\ref{sec:why-hard}
by exhibiting a concrete model
in which the response has suitable amplitude and
scale-dependence.
The model is compatible with current constraints on
the two- and three-point functions of $\zeta$,
and as we explain below it may
also compatible with constraints on the four-point function.
However,
its construction involves a number of arbitrary
choices.
We describe the model in~\S\ref{sec:tanh-model} before
summarizing the fine-tunings which are required.

A major advantage of working with a concrete model is that we
can compute its bispectrum in detail,
allowing the resulting shape and amplitude to be compared
with precision CMB constraints.
In~\S\ref{sec:gz-effect}
we discuss compatibility with constraints on the low
CMB multipoles $C_\ell$,
and in~\S\ref{sec:bsp-shape}
we estimate
the bispectrum amplitudes
$\hatfNLlocal$, $\hatfNLequi$ and $\hatfNLortho$
which would be measured by a Planck-like experiment
in this model.

\subsection{A step in $\eta$}
\label{sec:tanh-model}

To avoid difficulties with the $\zeta$ spectral index
we focus on models in which a large bispectrum is
generated during inflation, with no subsequent curvaton era,
as explained in~\S\ref{sec:no-const-eta}.
The discussion there
showed that we can not expect to obtain
a large amplitude if $\eta_\isomode$ is constant,
requiring the introduction of some feature
which allows $\eta_\isomode$ to evolve.
The next simplest possibility is to allow a step
which interpolates between two different constants.
To realize this we adopt the potential
\begin{equation}
\begin{split}
  V = V_0
  \bigg(
    1 + \frac{\eta_\phi}{2} \frac{\phi^2}{\Mp^2}
  \bigg)
  \bigg(
    1
    &
    \mbox{}
    + \frac{1}{2} \frac{\isomode^2}{\Mp^2}
    \bigg[
      \frac{\eta_2 - \eta_1}{2}
      \tanh \frac{\isomode-\isocrit}{\isograd}
      + \frac{\eta_1 + \eta_2}{2}
    \bigg]
    \\
    & \mbox{}
    - \frac{1}{2} \frac{\isocrit^2}{\Mp^2} \frac{\eta_2 - \eta_1}{2}
    \bigg[
      1 + \tanh \frac{\isomode - \isocrit}{\isograd}
    \bigg]
  \bigg) .
\end{split}
\label{eq:tanh-model}
\end{equation}
The step is centred at $\isomode = \isocrit$
and has characteristic width $\isograd$ in field units.
By making $\isograd$ small we can achieve a rapid transition.
Prior to the transition the effective $\eta_\isomode$
parameter is
$\eta_\isomode \approx \eta_1$.
After the transition we have
$\eta_\isomode \approx \eta_2$.
The scale $V_0$ should be chosen to match
the normalization of the $\zeta$ power spectrum.
In our numerical calculations
we take $V_0 \approx 10^{-14} \Mp^4$.

\begin{figure}
	\begin{center}
		\includegraphics[scale=1.0]{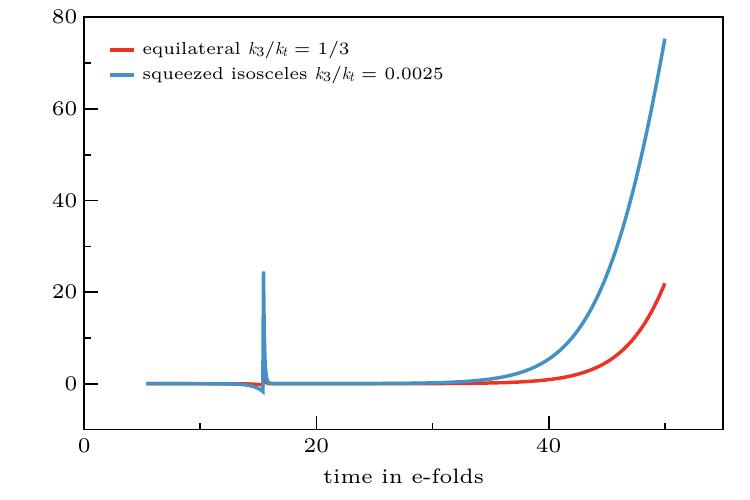}	
	\end{center}
	\caption{Time evolution of the reduced bispectrum $\fNL(k_1, k_2, k_3)$ in squeezed
	isosceles and equilateral configurations at the same scale $k_t$,
	chosen to be the value for which the average side $k_t/3$
	roughly corresponds to $\ell=1$.
	The spike at $\Ngrad \sim 15$ is the effect of the step
	in $\isomode$.
	The growth from $N \gtrsim 35$ represents the
	growth of the bispectrum amplitude
	predicted by~\eqref{eq:ridge-fNL-grow}
	as the initial kinetic energy imbalance is equalized.
	(Compare Fig.~\ref{fig:timeline}.)
	We have truncated the evolution at $N=50$.
	\label{fig:time-evolution-plot}}
\end{figure}

If we wish to work in a regime where $\delta \phi$ dominates
the $\zeta$ two-point function then we should choose
$\eta_\phi = \approx -0.02$
to give an acceptable spectral index.
We fix $\eta_1 = -0.25$
and $\eta_2 = -0.08$,
and take initial conditions
$\phi_\ast = 0.01 \Mp$ and
$\isomode_\ast = 8.94427 \times 10^{-8} \Mp$
at time $N=0$.
The diffusion regime exists for
roughly $|\isomode| < 4 \times 10^{-8} \Mp$,
making our initial conditions safe by
$\sim 3$ e-folds.
This is rather closer than one would like
in a realistic scenario, but our interest
is purely illustrative.
In any case it exemplifies the difficulty
of staying outside the diffusion regime---even
when features are introduced in the $\isomode$
potential.
Finally, taking the transition to occur
at $\isocrit = 3.445 \times 10^{-6} \Mp$
with width $\isograd = 10^{-10} \Mp$
gives an acceptable phenomenology
in which the step occurs at
a time $\Ngrad \sim 15$;
see Fig.~\ref{fig:time-evolution-plot},
which clearly demonstrates the need to tune
the time at which the inflationary phase exits.

In this paper we do not discuss an exit mechanism,
instead assuming that a choice can be found which
brings inflation to an end around $N=50$
while preserving the statistical properties of the field fluctuations.
Accurately accounting for effects associated with the end of inflation
and subsequent reheating \cite{Leung:2012ve,Meyers:2013gua,Elliston:2014zea}
would be a challenge for any attempt
to construct a realistic model describing the asymmetry.

\para{Computing the response}%
The large $\eta_\isomode$ prior to the step
and the rapidity of the transition
imply that we should not trust analytic
approximations for the correlation functions.
Instead we calculate estimates for the response
functions $\rho_\mu$
by combining Eq.~\eqref{eq:calculate-response}
with numerical computations
for the two- and
three-point functions of the model.
Our numerical method makes no use of the slow-roll
approximation, instead treating the dynamics exactly.
It also accounts for all quantum effects,
including deformation of the
wavefunctions due to mass terms,
mixing with the metric,
interference effects due to mode-coupling,
and off-diagonal correlations present
around the time of horizon exit.
However, in practice,
we find that these quantum effects are not very
significant for the model~\eqref{eq:tanh-model}.
The details of these simulations will be described
in forthcoming publications~\cite{long3pfpaper}.

\begin{figure}
  \begin{center}
    \includegraphics[scale=1.0]{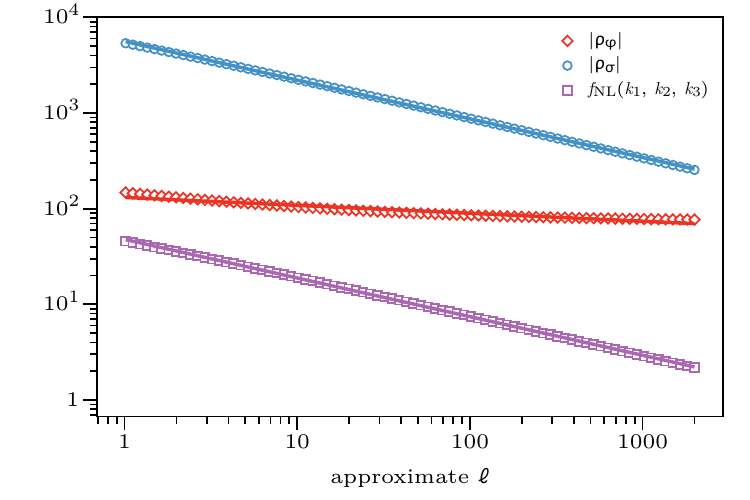}
    \includegraphics[scale=1.0]{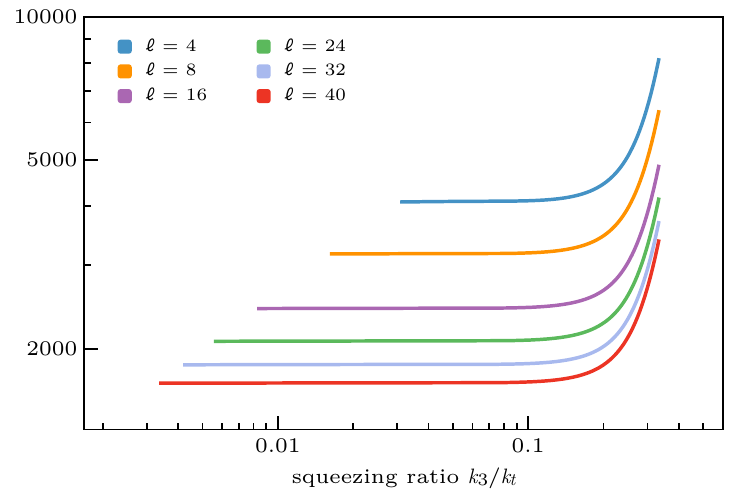}
  \end{center}
  \caption{Left panel:
  Absolute value
  of response functions $\rho_\phi$ and $\rho_\isomode$
  for the step model~\eqref{eq:tanh-model}.
  The responses are estimated
  using Eq.~\eqref{eq:calculate-response}
  on squeezed isosceles configurations where the
  long mode exits at time $N=0$
  and the approximate multipole $\ell = 14,000 k \, \text{Mpc}$
  corresponding
  to the short mode is plotted on the horizontal axis.
  We restrict to configurations for which
  the response can be estimated
  with $k_3/k_t < 0.1$
  in order to ensure that the response has become
  adequately independent of the long mode;
  see right panel.
  Also plotted is the reduced bispectrum amplitude
  $\fNL(k, k, k_3)$ measured on the same
  isosceles
  configurations used to estimate the
  response functions.
  The data points represent values extracted from
  our numerical method
  and the solid lines are power-law fits. \\
  Right panel: Variation in response functions
  with squeezing fraction
  $k_3/k_t$
  for different scales, measured by the approximate
  multipole $\ell$.
  The response functions become independent of $k_3$
  for squeezings $k_3 / k_t < 0.1$.
  \label{fig:response-plot}}
\end{figure}

\para{Numerical results}%
In the left panel of Fig.~\ref{fig:response-plot} we plot the response
functions $\rho_\phi(k)$ and $\rho_\isomode(k)$
as a function of the
approximate multipole $\ell = 14000 k \, \text{Mpc}$
corresponding to the wavenumber $k$.
The $\phi$ response function is reasonably close
to scale invariance,
but substantially smaller
than
the $\isomode$ response function which
exhibits strong scale-dependence.
It can be fit by the power law
\begin{equation}
  \rho_\isomode(k) \approx 5600 \left( \frac{k}{\kstar} \right)^{-0.405}
  \label{eq:tanh-response}
\end{equation}
where the scale $\kstar = 1/14,000 \, \text{Mpc}^{-1}$
corresponds to the multipole $\ell \approx 1$.
Eq.~\eqref{eq:tanh-response}
is roughly acceptable as a description of the scale
dependence of the modulation amplitude $A(k)$.%
    \footnote{We have verified that the response from
    the momentum perturbation
    $\d\delta\isomode/\d N$ is substantially smaller
    and can be neglected.}
The power-law for $\fNL(k, k, k_3)$ evaluated
on the same configurations (represented by the purple line
in the left panel of Fig.~\ref{fig:response-plot}) is
very close, with spectral index $-0.404$.
Both of these are close to the estimate
$\sim k^{2\eta_\isomode-2\eta_\phi} \sim k^{-0.36}$
obtained in Eq.~\eqref{eq:fNL-isosceles-slow-roll},
where we have approximated $n_s - 1 \approx 2\eta_\phi$,
and reproduce the conclusion of Eq.~\eqref{eq:fNL-asymmetry}
that $\fNL(k,k,k_3)$ and $\rho_\isomode(k)$
should scale similarly even if their amplitudes are different.
A very small amount of running is visible
in both $\rho_\isomode$ and $\fNL(k,k,k_3)$,
and fitting only to the region $1 \leq \ell \leq 60$ changes the spectral indices to
$-0.393$ and $-0.392$ respectively.

We have checked that one can obtain steeper power laws
if desired, by changing $\eta_1$
and modifying the initial conditions appropriately.
However because we are not seriously advocating this model
as an explanation of the anomaly, but using it only to
illustrate general properties,
we are content with the scaling~\eqref{eq:tanh-response}
which allows simple numerical values for $\eta_1$ and $\eta_2$.

These response functions are extracted from squeezed configurations,
and we have verified that the level of squeezing is sufficient
for each $\rho_\mu$ to become independent of the long mode.
This is demonstrated
in the right panel of Fig.~\ref{fig:response-plot}
which shows $\rho_\isomode$ as a function of the squeezing parameter
$k_3/k_t$.
In the left panel of Fig.~\ref{fig:response-plot}
we have included
response functions only
for values of $k$ which can be measured
from our numerics
on configurations with squeezing $k_3 / k_t < 0.1$.

\begin{figure}
  \begin{center}
    \includegraphics[scale=1.0]{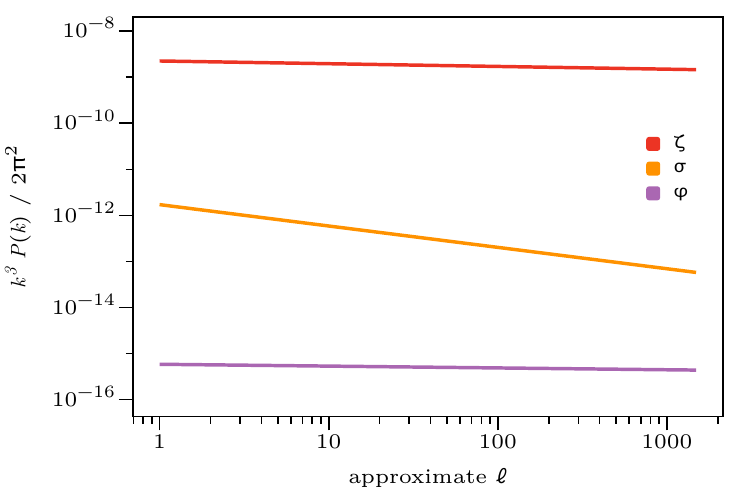}
    \includegraphics[scale=1.0]{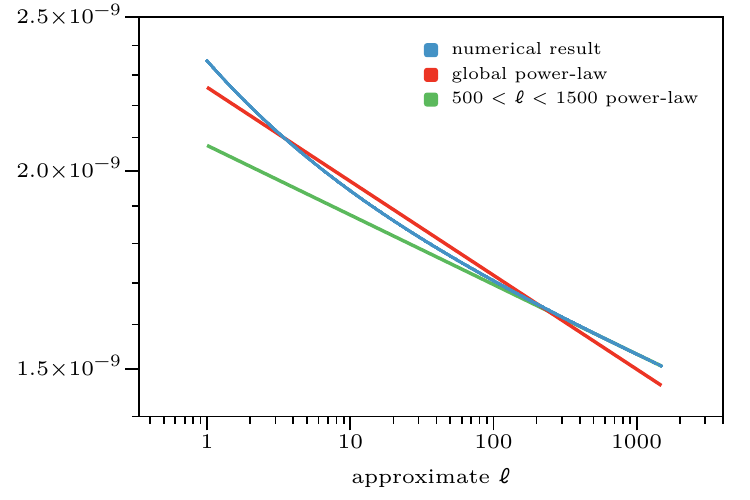}
  \end{center}
  \caption{Left panel: power spectra in the step model~\eqref{eq:tanh-model}, as
  a function of the approximate corresponding multipole
  $\ell \approx 14,000 k \, \text{Mpc}$. \\
  Right panel: Zoomed-in region showing only the $\zeta$
  power spectrum. The small curvature means that a
  global power-law,
  represented by the solid red line,
  is not a perfect fit; instead,
  some running is required.
  Fitting to only the region $500 < \ell < 1500$ (represented by
  the green line) gives a shallower
  power law.
  \label{fig:spectra-plot}}
\end{figure}

The power spectra for this model are shown in Fig.~\ref{fig:spectra-plot}.
On CMB scales
the $\zeta$ power spectrum is dominated by $\phi$,
but at small $\ell$
it begins to receive contributions
from $\isomode$ which generate running.
For the scales contributing to $\ell < 2000$
it can be fit by an approximate power law
of the form $k^{-0.06}$, which would correspond to
$n_s \sim 0.94$.
Fitting instead to the region $500 < \ell < 1500$
gives $n_s \approx 0.96$.
The $\isomode$ power spectrum
satisfies approximately
\begin{equation}
  	\dimlessP_\isomode(k) \approx 1.7 \times 10^{-12} \left( \frac{k}{\kstar} \right)^{-0.464} .
	\label{eq:concrete:Psigma}
\end{equation}

Assuming the larger-scale modes were generated during the same period
with large $|\eta_\isomode|$
we can estimate the required exceptionality $E$
as a function of the
quantity $\alpha$, introduced in Eq.~\eqref{eq:alpha-def}
to parametrize the physical scale of the modulating mode $\vectkL$,
\begin{equation}
  E(\alpha) \approx \frac{A(\kstar)}{\pi \alpha} \frac{1}{\rho_\isomode(k_2) \dimlessP_\isomode^{1/2}(\alpha \kstar)}
  \approx
  8.96 \alpha^{-0.768}
  = 8.96 \left( \frac{\kL}{\kstar} \right)^{-0.768} .
  \label{eq:tanh-exceptionality}
\end{equation}
In the final numerical estimate we have taken $A(\kstar) = 0.2$,
as suggested by the scale-dependent analysis by Aiola et al.~\cite{Aiola:2015rqa}.
It follows that we can achieve
exceptionalities in the desired range provided
$\kL$ is not too much smaller than $\kstar$.
As explained in~\S\ref{sec:perturbed-2pf}, this
is anyway required to control the term $C(k)$.

\subsection{The Grischuk--Zel'dovich effect}
\label{sec:gz-effect}

Having verified that a modulation with the correct amplitude
and scale
dependence can be
synthesized in this model, there are two observational checks
on its viability.
The first comes from
the requirement that the
amplitude of the modulating mode is not
so large that it would generate
unacceptable contributions to the low CMB multipoles.
This is the Grischuk--Zel'dovich effect.
The second constraint is that, for exceptionalities which pass the
Grischuk--Zel'dovich test, the required bispectrum amplitude
is compatible with Planck measurements
which are principally sensitive to squeezed configurations
associated with CMB scales.
In this section we pursue the Grischuk--Zel'dovich constraint,
leaving the bispectrum amplitude to~\S\ref{sec:bsp-shape}.

To compute the Grischuk--Zel'dovich effect,
we note that the modulating mode will make a
direct contribution to $\zeta$ which can be
estimated from~\eqref{eq:gauge-transformation}
after inverting the Fourier transform,
\begin{equation}
  \zeta(\vect{x}) \supseteq N_\isomode \delta\isomode(\vect{x})
  + \frac{1}{2} N_{\isomode\isomode} \delta\isomode(\vect{x})^2  
  + \cdots .
  \label{eq:zeta-gz}
\end{equation}
As in Eq.~\eqref{eq:gauge-transformation}
the fluctuations on both sides of this expression are to
be evaluated at the same time,
in contrast to Eq.~\eqref{eq:separate-universe-fields}.

The contribution
to CMB multipoles can be obtained
by combining~\eqref{eq:zeta-gz}
with the formula $\Theta(\hat{\vect{n}}) \approx \zeta(\xls \hat{\vect{n}}) / 5$
for the temperature anisotropy,
after
using~\eqref{eq:modulating-mode}
and
expressing the result as a spherical harmonic transform.
To do so we use
Rayleigh's formula for plane waves,
\begin{equation}  
  \exp(\im \vect{k} \cdot \vect{x})
  =
  4\pi
  \sum_{\ell m}
  \im^\ell
  j_\ell(kx)
  Y_{\ell m}^\ast(\hat{\vect{k}})
  Y_{\ell m}(\hat{\vect{x}}) ,
\end{equation}
where $j_\ell(x)$ is the spherical Bessel function of order $\ell$
and $Y_{\ell m}(\hat{\vect{n}})$ are spherical harmonics
oriented with respect to the polar axis $\hat{\vect{n}}$.
Using Rayleigh's formula to express the trigonometric functions
as spherical harmonic transforms, we conclude%
	\footnote{These estimates differ in detail compared with results previously reported
	in the literature~\cite{Erickcek:2008jp,Lyth:2013vha,Lyth:2014mga,Kanno:2013ohv,Kobayashi:2015qma},
	which replace the spherical Bessel functions with other numerical factors.
	The difference arises because
	$a_{10}$ and $a_{30}$ receive contributions from all odd powers
	of $\vectkL\cdot\vect{x}$
	whereas $a_{20}$ receives contributions from all even powers.
	Therefore any estimate
	requires an assumption about these higher order terms.
	The estimates
	made in
	Refs.~\cite{Erickcek:2008jp,Lyth:2013vha,Lyth:2014mga,Kanno:2013ohv,Kobayashi:2015qma}
	assumed a power series expansion in $\vectkL\cdot\vect{x}$
	which terminated at the quadratic or cubic level,
	so all higher-order terms were absent.
	In Eqs.~\eqref{eq:a10-gz}--\eqref{eq:a30-gz}
	we have assumed the higher-order terms come from expansion of~\eqref{eq:modulating-mode}.
	The difference between these estimates
	is already negligible for $\alpha \lesssim 0.1$.

	Our formula for the response was accurate only to $\Or(\kL^2)$,
	but this does not prevent us from using terms
	such as $(\vect{\kL}\cdot\vect{x})^2$
	or higher which combine $\vectkL$ with the same number
	of powers of $\vect{x}$.
	The $\Or(\kL^2)$ corrections would not involve
	$\vect{x}$
	and are therefore subleading compared to the
	angular terms retained in these estimates.
	
	Finally,
	the ansatz~\eqref{eq:modulating-mode} takes the long-wavelength perturbation to be
	a pure cosine at the time of interest, whereas in Ref.~\cite{Kobayashi:2015qma}
	it was taken to be a pure cosine at the time of horizon exit.
	Nonlinear evolution between horizon exit and the time of interest
	would then generate $\cos^2$ and higher contributions sourced by $\Gamma^\alpha_{ab}$
	and higher derivatives. These are not important for determining the
	asymmetry but would contribute to Eqs.~\eqref{eq:a10-gz}--\eqref{eq:a30-gz}.
	In both cases the pure cosine is just an ansatz and there will be further
	corrections which are being neglected, so we do not view this difference as material.}
\begin{subequations}
\begin{align}
  \label{eq:a10-gz}
  a_{10}
  & \supseteq
  - \frac{2 \sqrt{3\pi}}{5} E(\alpha) \dimlessP^{1/2}_\isomode(\kL)
    \Big(
      j_1(2\pi \alpha) N_\isomode \sin \vartheta
      + \frac{1}{4} E(\alpha) \dimlessP^{1/2}_\isomode(\kL) j_1(4\pi \alpha) N_{\isomode\isomode} \sin 2\vartheta
    \Big)
  \\
  \label{eq:a20-gz}
  a_{20}
  & \supseteq
  - \frac{2 \sqrt{5\pi}}{5} E(\alpha) \dimlessP^{1/2}_{\isomode}(\kL)
    \Big(
      j_2(2\pi\alpha) N_\isomode \cos \vartheta
      + \frac{1}{4} E(\alpha) \dimlessP^{1/2}_\isomode(\kL) j_2(4\pi \alpha) N_{\isomode\isomode} \cos 2 \vartheta
    \Big)
  \\
  \label{eq:a30-gz}
  a_{30}
  & \supseteq
  \frac{2 \sqrt{7\pi}}{5} E(\alpha) \dimlessP^{1/2}_{\isomode}(\kL)
    \Big(
      j_3(2\pi\alpha) N_\isomode \sin \vartheta
      + \frac{1}{4} E(\alpha) \dimlessP^{1/2}_\isomode(\kL) j_3(4\pi\alpha) N_{\isomode\isomode} \sin 2\vartheta
    \Big)
\end{align}
\end{subequations}

One can suppress some contributions to Eqs.~\eqref{eq:a10-gz}--\eqref{eq:a30-gz}
by tuning the phase $\vartheta$,
but this is unattractive
in a model which already requires significant tunings.
As in {\S}\ref{sec:perturbed-2pf}
we assume that the Earth lies at a typical point
where $\sin \vartheta \sim \cos \vartheta \sim \sin 2\vartheta \sim \cos 2\vartheta \sim \Or(1)$.
In the step model~\eqref{eq:tanh-model},
$N_\isomode$ settles down to a scale-independent
value $N_\isomode \approx -14.2568$ long after the transition,
and likewise for $N_{\isomode\isomode} \approx -8.07831$.
The conclusion is that the Grischuk--Zel'dovich contributions
to $a_{10}$, $a_{20}$ and $a_{30}$ can be written
\begin{subequations}
\begin{align}
  a_{10}
  & \supseteq
  2.04 \times 10^{-4} \frac{j_1(2\pi\alpha)}{\alpha} +
  3.38 \times 10^{-10} \frac{j_1(4\pi\alpha)}{\alpha^2}
  \\
  a_{20}
  & \supseteq
  2.64 \times 10^{-4} \frac{j_2(2\pi\alpha)}{\alpha} +
  4.37 \times 10^{-10} \frac{j_2(4\pi\alpha)}{\alpha^2}
  \\
  a_{30}
  & \supseteq
  3.13 \times 10^{-4} \frac{j_3(2\pi\alpha)}{\alpha} +
  5.17 \times 10^{-10} \frac{j_3(4\pi\alpha)}{\alpha^2} .
\end{align}
\end{subequations}

\begin{table}
  \begin{center}
    \small
	\heavyrulewidth=.08em
	\lightrulewidth=.05em
	\cmidrulewidth=.03em
	\belowrulesep=.65ex
	\belowbottomsep=0pt
	\aboverulesep=.4ex
	\abovetopsep=0pt
	\cmidrulesep=\doublerulesep
	\cmidrulekern=.5em
	\defaultaddspace=.5em
	\renewcommand{\arraystretch}{1.5}
        
    \rowcolors{2}{gray!25}{white}

    \begin{tabular}{suuuu}
      \toprule
      & \multicolumn{3}{c}{scale of modulating mode} & \multicolumn{1}{c}{observational limit} \\ \cmidrule{2-4}
      & \alpha=0.1 & \alpha=0.01 & \alpha=0.001 & \\
      a_{10} & 4.12 \times 10^{-4} & 4.28 \times 10^{-4} & 4.30 \times 10^{-4} & C_1^{1/2} \lesssim 10^{-3} \\
      a_{20} & 6.76 \times 10^{-5} & 6.95 \times 10^{-6} & 6.99 \times 10^{-7} & C_2^{1/2} \lesssim 5.9 \times 10^{-6} \\
      a_{30} & 7.22 \times 10^{-6} & 7.39 \times 10^{-8} & 7.48 \times 10^{-10} & C_3^{1/2} \lesssim 9.1 \times 10^{-6} \\
      E & 50.5 & 308 & 1800 & \\
      |C(\kstar)| & 0.32 & 3.2 & 32 & \\
      \bottomrule
    \end{tabular}
  \end{center}
  \caption{Grischuk--Zel'dovich contributions to
  $a_{10}$, $a_{20}$ and $a_{30}$ for the step model~\eqref{eq:tanh-model}
  at different values of $\alpha$, together with the exceptionality $E$
  and the monopolar amplitude modulation $|C(\kstar)|$.
  Limits on the observed quadruole and octupole $C_2$ and $C_3$
  have been given by Efstathiou~\cite{Efstathiou:2003tv},
  who reported $\Delta T_2^2 \lesssim 250 (\mu K)^2$
  and $\Delta T_3^2 \lesssim 1183 (\mu K)^2$
  with the real value expected to lie near these limits.
  These correspond to
  $C_2^{1/2} \lesssim 5.9 \times 10^{-6}$
  and
  $C_3^{1/2} \lesssim 9.1 \times 10^{-6}$.
  The dipole is comparatively unconstrained owing
  to uncertainties from the
  Earth's motion relative to the CMB rest frame,
  but the values $\sim 10^{-4}$ given here
  are acceptable because they are smaller than the observed value $\sim 10^{-3}$.
  \label{table:gz-contributions}}
\end{table}

Numerical values for $a_{10}$, $a_{20}$ and $a_{30}$, together with the
corresponding exceptionality $E(\alpha)$
and monopolar amplitude modulation $|C(\kstar)|$,
are shown in Table~\ref{table:gz-contributions}
for $\alpha = 0.1$, $0.01$ and $0.001$.
We also display observational limits
obtained from measurements
of the low $C_\ell$, where
\begin{equation}
  C_\ell = \frac{1}{2\ell + 1} \sum_m |a_{\ell m}|^2 .
\end{equation}
We should therfore expect $a_{20} < C_2^{1/2}$ and
$a_{30} < C_3^{1/2}$.
The limits used here match those
of Erickcek et al.~\cite{Erickcek:2008jp}.
Kanno et al.
and Lyth
used instead $C_2^{1/2} < 6.5 \times 10^{-6}$~\cite{Kanno:2013ohv,Lyth:2014mga}
and Kobayashi et al. used $C_2^{1/2} < 1.0 \times 10^{-5}$~\cite{Kobayashi:2015qma}.
In addition, Refs.~\cite{Lyth:2013vha,Lyth:2014mga}
used
the method described above to estimate
the contribution proportional to $N_{\isomode\isomode}$
(described as the `Erickcek--Kamionkowski--Carroll' effect),
but a different method to estimate
the contribution proportional to $N_\isomode$
(described as the `Grischuk--Zel'dovich' effect).
These different numerical choices led to $\Or(1)$ discrepancies between
the constraints on $\fNL$ reported in Refs.~\cite{Lyth:2013vha,Lyth:2014mga}
and Ref.~\cite{Kobayashi:2015qma}.
We follow Kobayashi et al. in treating the contributions
proportional to $N_\isomode$ and $N_{\isomode\isomode}$ in the same way.

From Table~\ref{table:gz-contributions}
we conclude that, in the specific model~\eqref{eq:tanh-model},
a value of $\alpha$
just a little smaller than $\alpha = 0.01$ should be acceptable,
requiring an enhancement factor a little larger than $E=300$
and perhaps a $1\%$ to $10\%$ tuning of $C(k)$.
The Grischuk--Zel'dovich contributions to $a_{20}$ and $a_{30}$
can be suppressed comfortably below the observed constraints
if we go as far as $\alpha = 0.001$ at the expense of
a much larger exceptionality $E \sim 2000$
and substantially more tuning in $C(k)$.
These values for $E$---but \emph{not} the tuning in $C(k)$, cf.
Eq.~\eqref{eq:Ck-alpha-Ak}---could be reduced
by generating a bispectrum of larger amplitude,
provided it remains compatible with the observational
constraints discussed in~\S\ref{sec:bsp-shape}.

\para{Comparison with earlier literature}%
If desired, Eqs.~\eqref{eq:a10-gz}--\eqref{eq:a30-gz}
can be rewritten in terms of $A$ and $\fNL$.
Focusing on the part of $\fNL$ generated by the superhorizon mode
and neglecting scale- and shape-dependence
of each quantity,
Refs.~\cite{Lyth:2013vha,Lyth:2014mga,Kanno:2013ohv,Kobayashi:2015qma}
found a relation of the form
\begin{equation}
	|a_{20}| \supseteq
	6.9\times 10^{-6}
	\times
	\frac{60}{|\fNL|}
	\left(
		\frac{A}{0.07}
	\right)^2
	|\beta(\alpha,\kL)|
	\label{eq:a20-A-fNL-basic}
\end{equation}
where $\beta$ is defined by
\begin{equation}
	\beta(\alpha,\kL)
	\approx
		\cos 2\vartheta
		+
		\frac{N_\isomode}{N_{\isomode\isomode}} \frac{1}{E(\alpha) \dimlessP_\isomode^{1/2}(\kL)}
		\cos \vartheta
\end{equation}
for $\alpha \ll 1$.
We have verified that a similar relation holds
with scale- and shape-information retained,
although because it is no more informative than~\eqref{eq:a20-A-fNL-basic}
we do not write it explicitly.

The quantity $\beta$ was
introduced
by
Kobayashi et al.~\cite{Kobayashi:2015qma}
and measures the relative contribution of the
$N_\isomode$ and $N_{\isomode\isomode}$
contributions in~\eqref{eq:a20-gz}.
In $\beta$ these translate to the terms
proportional to $\cos \vartheta$ and $\cos 2\vartheta$, respectively.
For the purposes of numerical estimates Kobayashi et al. assumed $|\beta| = \Or(1)$
which implies the $N_{\isomode\isomode}$ term is dominant.
In practice the $N_\isomode$ term is often more important
because it is enhanced by $\dimlessP_\isomode^{-1/2}$.
Without tuning $\vartheta$ this makes values in the range
$10$ to $10^3$ reasonable. 

Ignoring scale- and shape-dependence,
Eq.~\eqref{eq:a20-A-fNL-basic}
suggests that,
without unexpected
cancellation between Eq.~\eqref{eq:a20-A-fNL-basic}
and other contributions to $a_{20}$, 
we should expect $\fNL\gtrsim60$ even in the optimistic case $\beta\sim1$.
This led to a
discussion
in the literature
regarding compatibility of the model with observation,
since the amplitude of local-type contributions to the bispectrum
is now constrained to be substantially less than $60$.
In the next section, we will explicitly show that accounting for
the scale- and shape-dependence of the bispectrum allows us to
make $a_{20}$ sufficiently small and $A$ sufficiently large
without demanding an unacceptable amplitude,
even when we allow $\beta$ to be substantially larger than unity.

\subsection{The shape and amplitude of the bispectrum}
\label{sec:bsp-shape}

Finally we must check the amplitude of three-point correlations.
We have already observed that the reduced bispectrum
$\fNL(k_1, k_2, k_3)$ will run as a function of scale and squeezing,
and therefore will not match the `local' template used
to obtain the Planck2015 constraint $\fNLlocal = 0.8 \pm 5.0$~\cite{Ade:2015ava}.
In this section we study the shape of the bispectrum
generated by~\eqref{eq:tanh-model} in more detail.

\begin{figure}
  \begin{center}  
    \includegraphics[scale=1.0]{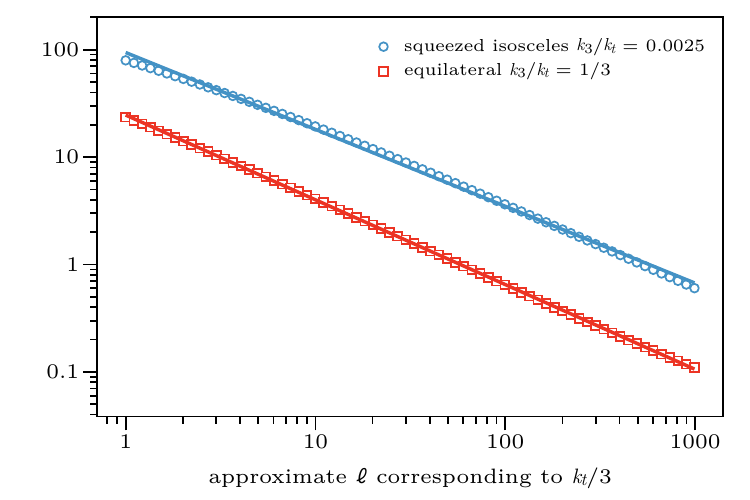}
    \includegraphics[scale=1.0]{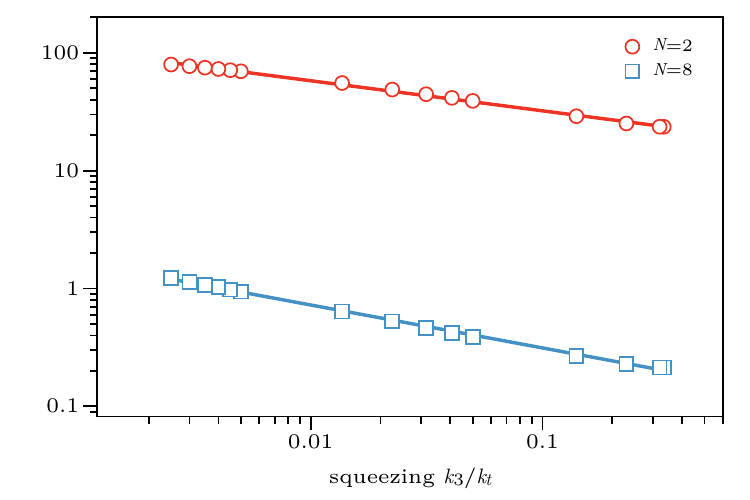}  
  \end{center}
  \caption{Dependence of the reduced bispectrum $\fNL(k_1, k_2, k_3)$ on scale and
  shape. The marked points are samples from our numerical results. \\
  Left panel: variation of $\fNL(k_1, k_2, k_3)$ with $k_t$
  (represented by the corresponding approximate multipole $\ell$)
  at fixed shape for
  two different shapes, (i) equilateral triangles, and (ii) squeezed isosceles triangles
  with $k_3/k_t=0.0025$. \\
  Right panel: variation of $\fNL(k_1, k_2, k_3)$ with $k_3/k_t$ at fixed scale.
  \label{fig:bispectrum-plot}}
\end{figure}

\para{Variation with scale}%
In Fig.~\ref{fig:bispectrum-plot}
we show the dependence of the bispectrum on $k_t$
for fixed shape,
and its dependence on squeezing $k_1/k_t$ at fixed scale.
The variation with $k_t$ at fixed shape can be fairly
well fit by a constant power law.
On equilateral triangles the amplitude is roughly
\begin{equation}
  \fNL = 24 \left( \frac{k_t/3}{\kstar} \right)^{-0.789} ,  
  \label{eq:fNL-kt-equilateral}
\end{equation}
and on squeezed triangles with $k_1/k_t = 0.0025$
we have
\begin{equation}
  \fNL = 94 \left( \frac{k_t/3}{\kstar} \right)^{-0.716} .
  \label{eq:fNL-kt-squeezed}  
\end{equation}
Bearing in mind the size of $\eta_\isomode$,
the scaling in~\eqref{eq:fNL-kt-equilateral}
is a reasonable match for our lowest-order
slow-roll prediction
$k_t^{4\eta_\isomode - 4 \eta_\phi}
\sim k_t^{-0.92}$
obtained in~\eqref{eq:fNL-kt-slow-roll}---%
and an even better match if the only large next-order
term is included to give
$k_t^{4\eta_\isomode + 4\eta_\isomode^2/3 -4 \eta_\phi}
\sim k_t^{-0.84}$~\cite{Byrnes:2009pe,Byrnes:2010ft},
within $6\%$ of the measured result.
(We have again
approximated $n_s-1 \approx 2 \eta_\phi$
as in~\S\ref{sec:tanh-model}.)
The accuracy of the slow-roll prediction is rather
striking,
and we have verified that similar accuracy
persists even for larger values of $|\eta_1|$.

Comparison of Eqs.~\eqref{eq:fNL-kt-equilateral} and~\eqref{eq:fNL-kt-squeezed}
shows that the $k_t$ dependence varies with shape,
in contradiction with the slow-roll
prediction~\eqref{eq:fNL-kt-slow-roll} which
depends only on the $\Gamma$-matrices and is shape-independent.
We interpret the shape dependence as a dominant contribution
from~\eqref{eq:fNL-kt-slow-roll} corrected by
a smaller shape-dependent contribution
from the $n$-point functions
$\langle \delta \phi^a \delta\phi^b \rangle$,
$\langle \delta \phi^a \delta\phi^b \delta\phi^c \rangle$
which were neglected in~\eqref{eq:fNL-kt-slow-roll}.
In this model the variation in these $n$-point functions
need not be small because of the large $\eta_\isomode$
when relevant scales were leaving the horizon.

Collecting Eqs.~\eqref{eq:tanh-response}, \eqref{eq:concrete:Psigma}
and~\eqref{eq:fNL-kt-equilateral}
yields the simple relations
\begin{equation}
	A(k)
	\sim
	\fNL(k,k,k_3)
	\sim
	\fNL^{1/2}(k,k,k)
	\sim
	\frac{\dimlessP_\isomode(k)}{\dimlessP(k)} ,
	\label{eq:special-relations}
\end{equation}
which should be interpreted as statements about the scaling behaviour
of each quantity as a function of $k$, with all other quantities such
as $k_3$ held fixed.
The first follows from Eq.~\eqref{eq:fNL-asymmetry}
and applies to any multi-source scenario
in which a single
source generates the bispectrum.
The second follows from
our assumption that the scale-dependence
is generated by a large $\eta_\isomode$
and follows from Eqs.~\eqref{eq:fNL-isosceles-slow-roll} and~\eqref{eq:fNL-kt-equilateral}.
Finally, the third
relation is another consequence of our assumption that
$\isomode$ dominates the bispectrum.
The same scalings~\eqref{eq:special-relations} will apply
to any scenario which
satisfies these criteria.
Notice that the asymmetry scales like $\fNL^{1/2}(k,k,k)$,
in contrast to the single-source case
for which the asymmetry is independent of the scaling
of the
$\isomode$ power spectrum and we instead have
\begin{equation}
	A(k)
	\sim
	\fNL(k,k,k)
	\quad
	\text{(single source)} .
\end{equation}

\para{Variation with shape}%
The right-hand panel of Fig.~\ref{fig:bispectrum-plot}
shows the variation with squeezing
$k_3/k_t$ on isosceles triangles at fixed scale.
There is an approximate fit to a constant power law,
but also some evidence for a change in the slope
between large and small $k_3/k_t$.
For the configuration whose average side
$k_t/3$ left the horizon at time $N=2$ we have, approximately,
\begin{equation}
  \fNL \approx 58 \left( \frac{k_3/k_t}{0.01} \right)^{-0.255} ,
  \label{eq:fNL-shape-large}
\end{equation}
and for the configuration whose average side
left the horizon at time $N=8$ we have,
approximately,
\begin{equation}
  \fNL \approx 0.73 \left( \frac{k_3/k_t}{0.01} \right)^{-0.365} .
  \label{eq:fNL-shape-small}   
\end{equation}

Eqs.~\eqref{eq:fNL-shape-large}
and~\eqref{eq:fNL-shape-small}
show reasonable agreement with the slow-roll
prediction~\eqref{eq:fNL-squeezing-slow-roll}.
The discrepancy is presumably accounted for
as above by
corrections from larger-than-slow-roll
scaling of
$\langle \delta\phi^a \delta\phi^b \rangle$,
$\langle \delta\phi^a \delta\phi^b \delta\phi^c \rangle$.

\para{Observational constraints}%
Fig.~\ref{fig:bispectrum-plot}
shows that the bispectrum amplitude is large on
some configurations but small on others.
The constraints reported by
the Planck collaboration
are limits on the amplitude of
scale-independent templates~\cite{Ade:2013ydc,Ade:2015ava}
averaged over many configurations,
and therefore as explained in~\S\ref{sec:introduction}
none of these can be related directly
to a bispectrum which runs significantly with scale.

To determine how the estimators for the
local, equilateral and orthogonal amplitudes
would respond
to the bispectrum produced by~\eqref{eq:tanh-model}
we construct a Fisher matrix estimate.
We numerically compute $\sim 5 \times 10^6$
bispectrum configurations for~\eqref{eq:tanh-model}
covering the
range from $\ell \sim 1$ to $\ell \sim 7000$
and use these to predict the
observed angular bispectrum $b_{\ell_1 \ell_2 \ell_3}$
up to $\ell \sim 2000$
using the method of Refs.~\cite{Fergusson:2009nv,Fergusson:2010gn,Regan:2013wwa,Regan:2013jua}
and
realistic estimates for the Planck beam and noise~\cite{Baumann:2008aq}.

\begin{figure}
	\begin{center}
		\includegraphics[scale=0.32,angle=0,trim= 0cm 4cm 0cm 0cm,clip=true]{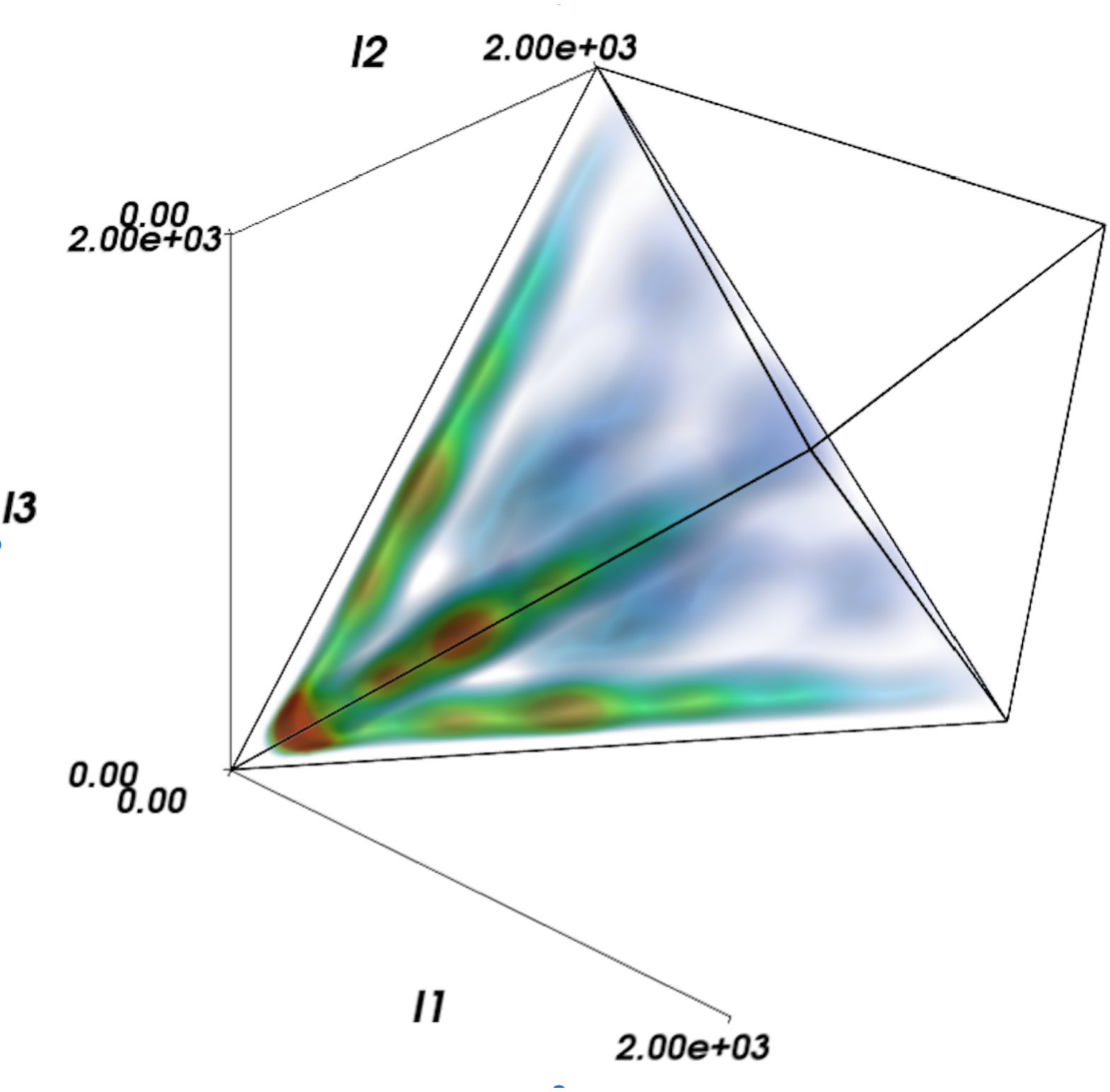}
		\includegraphics[scale=0.33,angle=0,trim= 1cm 4cm 0cm 2cm,clip=true]{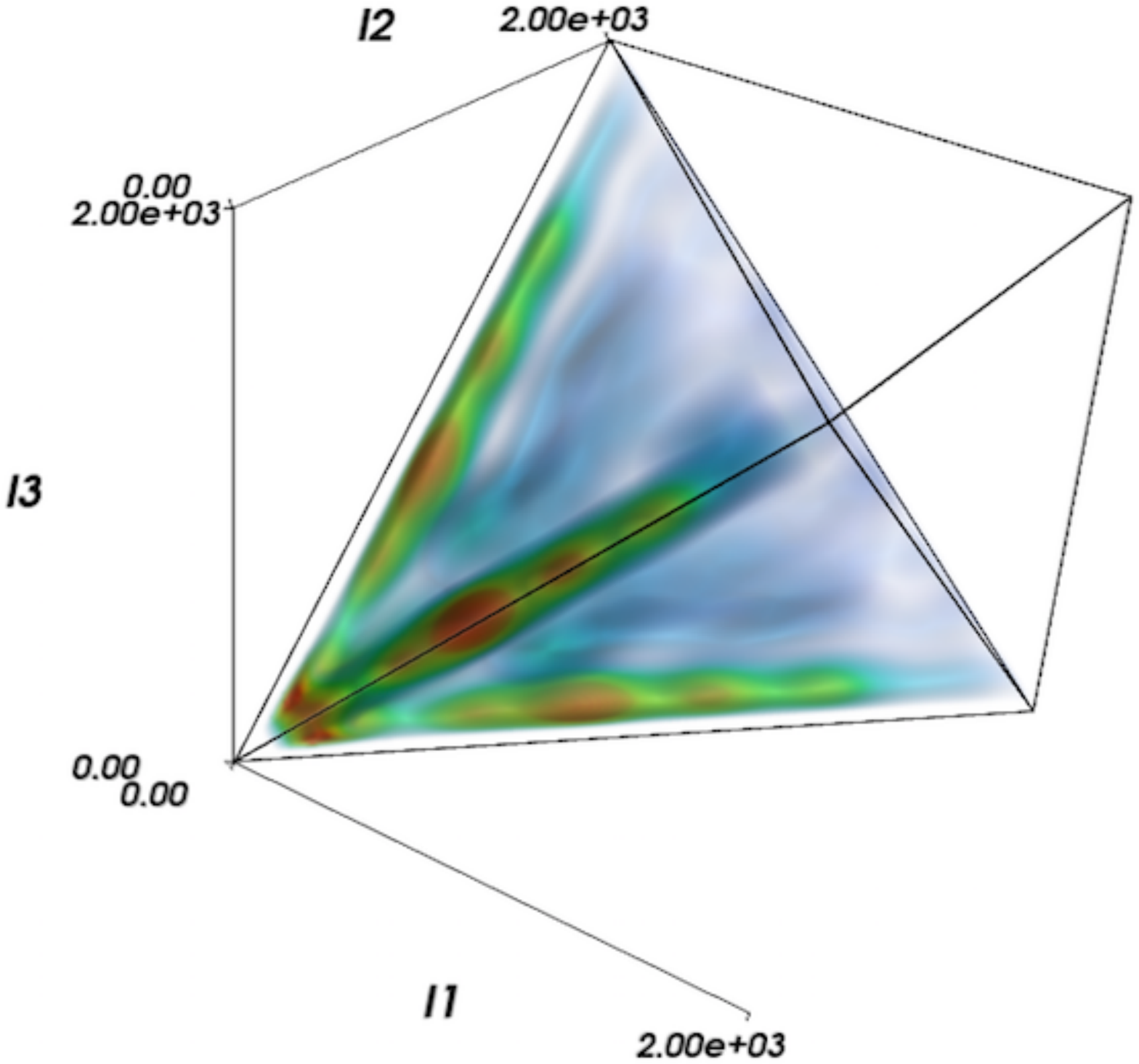}
	\end{center}
	\caption{Left: $\ell$-space bispectrum generated by the model~\eqref{eq:tanh-model}. \\
	   Right: $\ell$-space bispectrum generated by the local template. \\
	   The shapes are $95\%$ correlated even though the underlying $\vect{k}$-space
	   bispectra exhibit only $60\%$ correlation.
	   The $\ell$-space shapes are normalised as described in \cite{Ade:2015ava}.
	   Red regions represent positive values and blue regions representing negative values.
	   By comparison with the local template,
	   the red tilt can be seen qualitatively to enhance power at small $\ell$.
	\label{fig:bispec_shapes}}
\end{figure}

We find that the primordial bispectrum
$B(k_1, k_2, k_3)$ is roughly $60\%$ correlated
with the local template in $\vect{k}$-space.
Although the shape produced by~\eqref{eq:tanh-model}
is `local-like' in the sense of~\eqref{eq:ope}, the decorrelation can be attributed
to the strong running with shape and scale.
Despite the modest $\vect{k}$-space
correlation, the angular bispectrum
$b_{\ell_1 \ell_2 \ell_3}$ is
$95\%$ correlated
with the local template in $\ell$-space.
The difference is caused by redistribution
of power in the mapping from
$\vect{k}$ to $\ell$,
which 
reroutes structure towards modestly
squeezed $\ell$-configurations by mixing contributions
from the low-$k$ regime.
This increases
overlap with the local shape,
although there is still
marginally enhanced power on
large scales; see Fig.~\ref{fig:bispec_shapes}.

In $\ell$-space
our bispectrum correlates at $98\%$ with a
local template normalized
as in Eqs.~\eqref{eq:fNL-kt-equilateral}--\eqref{eq:fNL-kt-squeezed},
but with
amplitude scaling as the power-law $k_t^{-0.7}$
and no dependence on the squeezing $k_i/k_t$.
This template is
a good match for our bispectrum in $\ell$-space,
and a better approximation than the pure local template.

For scale-independent bispectrum shapes it usually happens
that the $\vect{k}$-space correlation is a good predictor for the
$\ell$-space correlation.
In our example this is not true
and the $\vect{k}$-space correlation would produce
a misleading result.
We believe this to be a fairly
general feature of scale-dependent
shapes, and if so it will imply that
a running bispectrum shape intended to explain the
asymmetry must be projected into $\ell$-space
before robust conclusions can be extracted
regarding its observational viability.

We find that the amplitudes which would be measured
for a bispectrum generated by~\eqref{eq:tanh-model}
are order unity.
We obtain
\begin{equation}
  \hatfNLlocal = 0.25
  ,
  \quad
  \hatfNLequi = 0.6
  ,
  \quad
  \hatfNLortho = -1.0
  .  
  \label{eq:fNL-estimator-predictions}
\end{equation}
These amplitudes
can be regarded as weighted averages of $\fNL(k_1, k_2, k_3)$
over many configurations.
Since Fig.~\ref{fig:bispectrum-plot}
shows that some configurations reach amplitudes
of order $\Or(100)$,
the $\Or(1)$ values reported in~\eqref{eq:fNL-estimator-predictions}
imply that,
on the configurations which contribute most signal-to-noise
to the estimators
$\hatfNLlocal$, $\hatfNLequi$ and $\hatfNLortho$,
the amplitude $\fNL(k_1, k_2, k_3)$ has already run to
small values;
in fact,
the amplitude of $\hatfNLlocal$ agrees well
with the value which would be inferred
from~\eqref{eq:fNL-kt-equilateral} evaluated
at the Planck pivot scale $0.05h \, \text{Mpc}^{-1}$
(corresponding approximately to $\ell\sim700$).
These numbers should be compared to the Planck2013
temperature-only%
  \footnote{We quote the temperature-only constraints
  because our analysis does not include polarization data.}
constraints~\cite{Ade:2013ydc}
\begin{equation}
  \hatfNLlocal = 2.5 \pm 5.7
  ,
  \quad
  \hatfNLequi = -16 \pm 70 
  ,
  \quad
  \hatfNLortho = -34 \pm 33 .
\end{equation}
We conclude that the bispectrum amplitude generated by
this model is well within the observational limits.

Eq.~\eqref{eq:fNL-estimator-predictions}
appears to constrast with the much
larger estimates
appearing in
Refs.~\cite{Lyth:2013vha,Lyth:2014mga,Kobayashi:2015qma,Adhikari:2015yya}.
However,
because our model still
requires $\fNL(k_1, k_2, k_3)$ to be large on the configurations
responsible for determining $A(k)$, our analysis does not disagree
with the qualitative conclusions in these papers.
Eqs.~\eqref{eq:fNL-estimator-predictions}
are predictions for what would be observed in a realistic
experiment, and this prediction is possible only because
we have a concrete model,
enabling the bispectrum to be accurately computed.

The numerical bispectrum we have used is strictly
valid only for the model of Eq.~\eqref{eq:tanh-model},
but in practice we believe it will be a good
proxy for the bispectrum generated
in any model which uses a large $\eta_\isomode$.
If so, then after suitable
rescaling Eq.~\eqref{eq:fNL-estimator-predictions} may be used to
estimate the amplitudes produced in any model
designed to generate the asymmetry by this mechanism.

Because our estimates~\eqref{eq:fNL-estimator-predictions}
are so small there is
enough headroom,
if desired,
to increase the amplitude of the bispectrum
and decrease the exceptionality $E$.
Based on the $95\%$ correlation with the local
template in $\ell$-space
we expect
it is possible to increase the bispectrum amplitude
by a factor of roughly $\sim 50$
while staying within the $2\sigma$
error bar.
If this were done it would allow
$\fNL(k_1, k_2, k_3)$ to have
an amplitude as large as $\Or(10^3)$
on the configurations which determine $A(k)$
while still satisfying observational constraints.

\para{Trispectrum}%
We do not study the trispectrum in detail, because to do so accurately
would require numerical calculations of the four-point functions which
have not yet been developed.
Instead we use Eq.~\eqref{eq:tauNL-estimate}
to estimate the amplitude of $\tauNL$,
which should give fair results for both tetrahedral
and collapsed configurations.
However, because the scale- and shape-dependence
of the $\tauNL$ shape is rather strong,
comparison with the upper bound $\tauNL < 2800$
reported by Planck at $95\%$-confidence~\cite{Ade:2015ava}
is uncertain.
A trustworthy estimate should take into account which configurations
contribute the largest signal-to-noise.
We provide numerical values for qualitative guidance only.

We find that the amplitude is strongly scale-dependent,
becoming very large on long scales
but running to smaller values on short scales.
We estimate that it has a rough scale-dependence $\propto k^{-0.7}$,
with amplitude running from $\tauNL \sim 30,000$ on the
scale $\ell \approx 1$ to $\tauNL \sim 400$ on the scale $\ell \approx 500$.
Bearing in mind that,
as for the bispectrum,
the signal-to-noise for
the estimator $\hattauNL$
may receive its largest contribution from configurations
at modest or high $\ell$,
these numbers suggest that $\tauNL$
is not so large that the model is obviously unacceptable.
However, this should be confirmed by a more accurate analysis.

\section{Conclusions}
\label{sec:conclusions}

In this section we collect our conclusions.

\para{Technical results}%
Our principal theoretical results are
Eqs.~\eqref{eq:calculate-response}
for the response function $\rho_\mu$ in a `local-like' model,
and~\eqref{eq:zeta-response-single-source}
for the response to a long-wavelength $\zeta$ perturbation
in the special case of a single-source model.
These results extend the analyses given in
Refs.~\cite{Lyth:2013vha,Lyth:2014mga,Kobayashi:2015qma,Kenton:2015jga}
which assumed a single-source model
for which slow-roll was a good approximation
while relevant scales were leaving the horizon,
and used the separate universe approximation
to estimate biasing of the short-wavelength
power spectrum.
Our key tools are the operator product expansion
and linear response theory, of the kind used widely
in applications of field theory to condensed matter.
Our method does not invoke the slow-roll
approximation and applies
to `local-like' models where the long-wavelength response of
each two-point function is dominated by
the operator $\delta\phi^\mu$.
It could be easily extended to obtain the response
of the two-point function to any local operator
of interest.

Specializing to the single-source, slow-roll
case we
reproduce
the formulae of
Lyth~\cite{Lyth:2013vha,Lyth:2014mga},
Namjoo et al~\cite{Namjoo:2013fka,Abolhasani:2013vaa,Namjoo:2014nra}
and Kobayashi, Cort\^{e}s \& Liddle~\cite{Kobayashi:2015qma}.
We verify the statement made in
Ref.~\cite{Lyth:2014mga},
that in these models
it is the reduced bispectrum on equilateral configurations
which controls the response of the two-point function
to biasing.
This is a special case of the more general result~\eqref{eq:fNL-asymmetry}
that if a single source dominates the bispectrum
then the asymmetry scales with $k$ like
$\fNL(k,k,k_3)$ at fixed $k_3 \ll k$.
The special feature of truly single-source models
is that $\fNL(k,k,k_3)$ is independent of $k_3$,
making the amplitude the same as the equilateral
configuration
$\fNL(k,k,k)$.

If more than one field contributes to the bispectrum then
Eq.~\eqref{eq:calculate-response}
shows that it is a combination of suitable response
functions $\rho_\mu$ rather than the reduced bispectrum which
will determine the response.
These response functions are determined from
squeezed isosceles configurations of the
mixed three-point function
$\langle \delta \phi^\alpha \zeta \zeta \rangle$.
However,
because
the linear response calculation for local-like models
predicts that
each $n$-point function
responds to all long-wavelength modes in the same
way, the response does not depend on the squeezing ratio $k_3/k_t$.
Nevertheless this does not mean it is related in any simple way
to
the bispectrum amplitude on equilateral configurations, although in
some models that will be the case.

In~\S\ref{sec:scale-dependence}
we have developed a formalism to compute the scale-
and shape-dependence of each $n$-point function
without invoking a perturbative expansion
of the `separate universe coefficients'
$\Gamma^\alpha_a$, $\Gamma^\alpha_{ab}$.

\para{Model building constraints}%
Even if we know how to compute the response it is still necessary
to construct a model.
In~\S\ref{sec:why-hard} we have described in
general terms why this is difficult.
If one generates scale-dependence using
a large $\eta$-parameter
associated with an isocurvature field $\isomode$
then
the potential cannot be featureless,
and it appears unavoidable that this introduces fine-tuning.
Also, the generic result is
contamination of the $\zeta$
spectral index
if $\isomode$ contributes to the
$\zeta$ two-point function,
or an enhanced trispectrum amplitude if it does not.
In either case it may also be necessary to tune
the time and mechanism by which inflation ends.
If one instead generates scale-dependence using
a large $\xi$-parameter
while keeping $\eta$ small
then the time dependence of $\xi$ must be tuned
to give an approximate power-law.
We have not succeeded in constructing an example
model of this type because
typically a large $\xi$ sources a large $\eta$
within a few e-folds,
and it is not clear whether this problem can be
overcome.%
	\footnote{It can be overcome if the effective
	$\xi$ is oscillatory, as in the model
	of Enqvist et al.~\cite{Enqvist:2014nsa}.
	Then $\eta$ remains small due to cancellation
	of opposite-sign contributions from
	$\xi$.
	However the oscillations lead
	to oscillatory effects in the bispectrum,
	meaning
	it is still not easy to manufacture
	an approximate power-law over a sufficient
	number of e-folds.}
	
To exemplify these difficulties we have
constructed an explicit model
giving an acceptable fit to present
constraints on the $\zeta$ two-, three- and
four-point functions, and avoiding
(if only marginally)
pathologies such as a quantum diffusion regime.
This model has a large $\eta$ parameter for the
isocurvature field $\isomode$,
and therefore the slow-roll approximation is
not obviously acceptable.
Instead, to obtain accurate
predictions,
we have used a numerical method to
estimate the two- and three-point functions
and the response functions.
We
find (perhaps surprisingly)
that the slow-roll estimates continue to apply.

To determine whether the bispectrum amplitude
is compatible with recent constraints from Planck
we compute the angular bispectrum
$b_{\ell_1 \ell_2 \ell_3}$
up to $\ell \sim 2000$
and obtain the response of
the local, equilateral and orthogonal estimators
$\hatfNLlocal$, $\hatfNLequi$ and $\hatfNLortho$.
We conclude that these are all order unity.
Despite the large
reduced bispectrum
$\fNL(k_1, k_2, k_3)$ on those configurations
responsible for
the asymmetry $A(k)$,
this shows that the model is comfortably compatible
with present-day constraints;
indeed, it is even possible to increase
the bispectrum amplitude if
we wish to decrease the required enhancement factor $E$.

\para{Phenomenology}%
Our numerical computations use the precise bispectrum generated
by the step-like model~\eqref{eq:tanh-model}.
However the details of the bispectrum
do not strongly depend on the model;
the close correspondence between the generic
estimates obtained in~\S\ref{sec:scale-dependence}
and those measured from our numerics
show that the bispectrum is mostly determined by the
large value for $\eta_\isomode$.
Neither the scale- or shape-dependence is
influenced by the $\tanh$ step,
which is only important to obtain a suitable amplitude.
Therefore the results reported in~\S\ref{sec:tanh-model}---%
especially the estimates for
$\hatfNLlocal$, $\hatfNLequi$ and $\hatfNLortho$---%
have a wider significance for models which attempt to
explain the
asymmetry using a bispectrum of this type.
In particular, our results could be used
to estimate $\hatfNLlocal$, $\hatfNLequi$ and $\hatfNLortho$
for such models by suitable rescaling.

In addition, we have clarified the scaling properties of $A(k)$
in different scenarios (\S\ref{sec:bsp-shape}). 
In truly single-source scenarios
it is already known that $A(k) \sim \fNL(k,k,k)$~\cite{Lyth:2013vha,Lyth:2014mga,Kobayashi:2015qma}.
In a more general class of scenarios
where a single source dominates the bispectrum
(but not necessarily the two-point function) we have shown
that $A(k) \sim \fNL(k,k,k_3)$ at fixed $k_3$.
If the field responsible for sourcing the bispectrum does
not dominate $\dimlessP(k)$ then
when rewritten in terms of the equilateral
amplitude $\fNL(k,k,k)$ the
dependence of $\fNL(k,k,k_3)$ on $k_3$
changes the scaling law to
$A(k) \sim \fNL^{1/2}(k,k,k)$.
In this sense the two scenarios are strikingly
different, rather than one being a perturbative
refinement of the other.

Using the operator product expansion
we provide the formulae~\eqref{eq:fNL-asymmetry}
and~\eqref{eq:tauNL-asymmetry}
which
relate the asymmetry amplitude $A(k)$
to $\fNL(k_1, k_2, k_3)$
on squeezed
configurations,
and $\tauNL(\vect{k}_1, \vect{k}_2, \vect{k}_3, \vect{k}_4)$
on collapsed configurations.
These relations
are model independent, assuming only that a single field
dominates the bispectrum and trispectrum respectively, and apply
for any bispectrum shape controlled by the local
part of the OPE.
Together they predict an enhanced $\tauNL$ amplitude
whenever the $\isomode$ field does not contribute
to the $\zeta$ two-point function.

\para{Discussion}%
In our opinion none of the early-universe
scenarios which have been proposed to date are compelling.
Whether we are forced to take them seriously may become
clearer once polarization data become available,
which will provide an independent
probe of the amplitude of long-wavelength modes.

If the asymmetry is not a statistical accident,
and is to be explained by using a bispectrum
to
couple long- and short-scale
modes,
then it has been understood for a long time
that a single-field inflationary model is not viable~\cite{Erickcek:2008sm}.
In this paper we
additionally
argue that
although multiple-field models satisfying the required
observational criteria may exist, they
typically require multiple
independent fine-tunings
including at least some of the following.

\begin{enumerate}

	\item The existence of a long-wavelength mode enhanced by an
	exceptionality $E\gtrsim 10$ compared to the na\"{\i}ve
	estimate from power-law scaling.
	In the absence of
	new physics to explain its amplitude this would correspond to a
	$10\sigma$ fluctuation or more, and is a poor explanation of a $3\sigma$ anomaly.
	If there is new physics which naturally makes the exceptionality 
	large---for example, in the scenarios of Refs.~\cite{Liddle:2013czu,Mazumdar:2013yta}---%
	then care must be taken to include the effect of many long-wavelength modes
	with similar amplitude~\cite{Schmidt:2012ky,Adhikari:2015yya}.
	This may perhaps lead to a large
	quadrupolar modulation of the power spectrum.%
		\footnote{Quadrupolar modulation would correspond to the
		next-order term in~\eqref{eq:twopf-modulation},
		giving an angular dependence of the form $(\hat{\vect{p}}\cdot\hat{\vect{n}})^2$.}
		
	\item A tuning of the Taylor coefficient of order
	$(\vectkL\cdot\vect{x})^0$
	with respect to that of order $(\vectkL\cdot\vect{x})^1$
	for the long-wavelength mode.
	For small $\alpha$ this is required to prevent $C(k) 
	\propto A(k)/\alpha$
	growing too large, as described below Eq.~\eqref{eq:Ck-alpha-Ak},
	and generating an unwanted scale-dependent, monopolar modulation of power.
	This tuning is independent of the exceptionality $E$
	and the amplitude of the bispectrum.

	\item A large effective mass, corresponding to a large $\eta_{\isomode}$, or large
	self-interaction of the scalar field $\isomode$ which generates the asymmetry.
	In both cases the initial value of $\isomode$
	requires significant fine-tuning to generate a bispectrum of sufficient amplitude.

	\begin{itemize}

		\item \para{Large $\eta_\isomode$}%
		In this case the power spectrum must predominantly be generated by the
		inflaton field $\phi$ in order to preserve the near scale-invariance of
		the $\zeta$ power spectrum. We have shown that, if this is achieved by
		keeping $R \equiv
		\dimlessP_{\zeta_\sigma} / \dimlessP_{\zeta} \lesssim 0.1$ throughout the evolution, then
		$\eta_\isomode$ cannot
		be constant
		without trespassing on the diffusion region near a hilltop of the potential.
		
		Although one can construct models which avoid this constraint by introducing
		features, such as
		the step-like $\tanh$ model in~\S\ref{sec:working-model},
		the location of the feature represents another fine-tuning.
		In addition it adds complexity to a model which is already
		complicated.
		Even if this can be done successfully the amplitude of the
		$\tauNL$ trispectrum shape is typically enhanced
		above its single-source value $\sim \fNL^2$.

		Because of the strong scale-dependence of the $\tauNL$ amplitude
		we are not yet able to determine conclusively whether
		the model of~\S\ref{sec:working-model}
		is ruled out by observation.

		\item \para{Large self-interaction}%
		In this case
		the large self-interaction will typically lead to a large effective mass
		as $\isomode$ evolves, potentially spoiling scale-invariance of the $\zeta$
		power spectrum. It also generates a large contribution to the
		amplitude of the $\gNL$ trispectrum shape.
		If $\xi$ is constant then this amplitude may be close to scale invariant
		(unlike $\fNL$ or $\tauNL$)
		and therefore in conflict with observation;
		alternatively,
		if $\xi$ evolves in such a way
		that it produces an approximate power-law bispectrum
		the situation is less clear.
		An example of this case was studied in
		Ref.~\cite{Byrnes:2015asa}, where it was demonstrated that
		the quadrupolar modulation of the power spectrum is typically
		much too large. 

	\end{itemize}

	\item The long-wavelength mode must respect
	observational constraints on $C_\ell$ for low $\ell$.
	Although we have not done so here, other authors have
	invoked fine-tuning of our position on the long wavelength mode to help
	evade these constraints~\cite{Kobayashi:2015qma}.
	Strategies of this sort are possible but unattractive,
	and indeed
	it is not clear whether
	they remain viable if enough $\ell$-modes are considered.
	Alternatively,
	the Grischuk--Zel'dovich contributions to $C_2$ and $C_3$
	can be supressed
	by increasing the wavelength of the large-scale
	mode, but that simultaneously
	increases the tuning required in $|C(k)|$.
	
\end{enumerate}

\noindent
All these challenges can be understood as
manifestations of the difficulty
in constructing a $\sim 20\%$ modulation of the power spectrum amplitude on large scales
while maintaining consistency with
observational constraints on the smallness of
the bi- and trispectrum, and
quadrupolar modulation of the power spectrum.
Even worse, the largest CMB multipoles are actually observed to be suppressed
whereas the existence of a large amplitude long-wavelength fluctuation
would naturally be expected to enhance them. It is challenging to construct any
model which seeks to explain these conflicting demands without
emerging as more unlikely than the hemispherical asymmetry itself.

\section*{Acknowledgements}
DS acknowledges support from the Science and Technology
Facilities Council [grant number ST/L000652/1].
CTB is a Royal Society University Research Fellow.
The research leading to these results has received funding from
the European Research Council under the European Union's
Seventh Framework Programme (FP/2007--2013) / ERC Grant
Agreement No. [308082].
This work was supported in part by
National Science Foundation Grant No. PHYS-1066293
and the hospitality of the Aspen Center for Physics.

Some numerical work was undertaken on the COSMOS Shared Memory system at
DAMTP, University of Cambridge, operated on behalf of the STFC DiRAC HPC Facility.
This equipment is funded by BIS National E-infrastructure
capital grant ST/J005673/1 and STFC grants ST/H008586/1, ST/K00333X/1.

We would like to thank
Shaun Hotchkiss,
Zac Kenton,
Takeshi Kobayashi,
Antony Lewis,
David Mulryne,
Marjorie Schillo
and
Sarah Shandera
for helpful discussions.

\para{Data availability statement}%
Please contact the authors to obtain the bispectrum for the
step model~\eqref{eq:tanh-model},
which was used to estimate the responses~\eqref{eq:fNL-estimator-predictions}.

\bibliography{refs}
\end{document}